\documentclass[aps,pra,twocolumn]{revtex4-2}
\usepackage{graphicx}
\usepackage{amsmath,upgreek}
\usepackage{physics}
\usepackage{hyperref}
\usepackage{times}
\usepackage{bbm}

\hypersetup{
    colorlinks=true,
    linkcolor=blue,
    filecolor=black,      
    urlcolor=blue,
    allcolors=blue,
}

\begin{document}

\title{Efficient, High-Fidelity Single-Photon Switch Based on Waveguide-Coupled Cavities}
\author{Mateusz Duda}\email{mduda1@sheffield.ac.uk}
\author{Luke Brunswick}
\author{Luke R. Wilson}
\author{Pieter Kok}\email{p.kok@sheffield.ac.uk}
\affiliation{Department of Physics and Astronomy, The University of Sheffield, Sheffield S3 7RH, United Kingdom}
\date{\today}


\begin{abstract}
We demonstrate theoretically that waveguide-coupled cavities with embedded two-level emitters can act as a highly efficient, high-fidelity single-photon switch. The photon switch is an optical router triggered by a classical signal --- the propagation direction of single input photons in the waveguide is controlled by changing the emitter-cavity coupling parameters in situ, for example using applied fields. The switch reflects photons in the weak emitter-cavity coupling regime and transmits photons in the strong coupling regime. By calculating transmission and reflection spectra using the input-output formalism of quantum optics and the transfer matrix approach, we obtain the fidelity and efficiency of the switch with a single-photon input in both regimes. We find that a single waveguide-coupled cavity can route input photon wave packets with near-unity efficiency and fidelity if the wave packet width is smaller than the cavity mode linewidth. We also find that using multiple waveguide-coupled cavities increases the switching bandwidth, allowing wider wave packets to be routed with high efficiency and fidelity. For example, an array of three waveguide-coupled cavities can reflect an input Gaussian wave packet with a full width at half-maximum of $1$~nm (corresponding to a few-picosecond pulse) with an efficiency ${E_r = 96.4\%}$ and a fidelity ${F_r = 97.7\%}$, or transmit the wave packet with an efficiency ${E_t = 99.7\%}$ and a fidelity ${F_t = 99.8\%}$. Such efficient, high-fidelity single-photon routing is essential for scalable photonic quantum technologies.
\end{abstract}

\maketitle


\section{Introduction}\label{sec:intro}

Creating large-scale, distributed quantum networks for technologies such as computing, communication, sensing, and metrology requires precise control over single photons~\cite{Kimble2008, OBrien2009, Northup2014, Slussarenko2019}. A fundamental building block of photonic quantum technologies is therefore a device that can deterministically and faithfully route a single photon within a network, i.e., a single-photon switch with high efficiency and fidelity. Photon routing based on linear optics alone is probabilistic~\cite{Lemr2013, Yuan2015, Bartkiewicz2018}, and there have been many proposals of single-photon switches that utilise light-matter interactions for deterministic operation. Previous theoretical studies of such photon routers involve emitter-waveguide systems~\cite{Chang2007, Li2015, Poudyal2020}, single resonators with~\cite{Bermel2006, Xia2013, Luo2023} and without~\cite{Agarwal2012, Yan2015, Gao2018} coupled emitters, and more complex cavity-based structures~\cite{Yan2016_2, Yang2018, Zhu2019_2}, including arrays of many coupled cavities~\cite{Zhou2008, Zhou2013, Lu2014}. Experimental platforms proposed for photon switching include superconducting circuits with transmon qubits~\cite{Shen2005_2, Hoi2011, Wang2021, Rinaldi2024}, semiconductors with embedded quantum dots~\cite{Sun2018, Papon2019, Munoz-Matutano2020}, atomic ensembles~\cite{Dawes2005, Bajcsy2009, Baur2014}, and single emitters such as nitrogen-vacancy centres~\cite{Cao2017} or atoms~\cite{Aoki2009, OShea2013, Shomroni2014} coupled to a microresonator. 

Coupled-cavity arrays (CCAs) are a particularly promising platform, as they exhibit quantum many-body phenomena that can be exploited in various photonic quantum technologies~\cite{Hartmann2008, Tomadin2010}. This includes entanglement generation~\cite{Bostelmann2023, Liew2012, Liew2013, Angelakis2010}, cluster state preparation~\cite{Sun2016}, and many-body phase transitions~\cite{Hartmann2006}. In many proposed CCAs, photon propagation is mediated by evanescent field coupling, which requires the electromagnetic fields of neighbouring cavities to overlap spatially, spectrally, and in $k$-space~\cite{Afzal2019}. Spatial overlap between cavity modes places significant constraints on the geometry of a CCA, most notably that the cavity separations must be on the wavelength scale~\cite{Altug2004}. This makes it challenging to address and manipulate the properties of each cavity individually, and hence to precisely control the propagation of photons and overcome fabrication disorder~\cite{Barclay2006, Notomi2008, Heuser2018}.

\begin{figure}[b!]
    \includegraphics[width = \columnwidth]{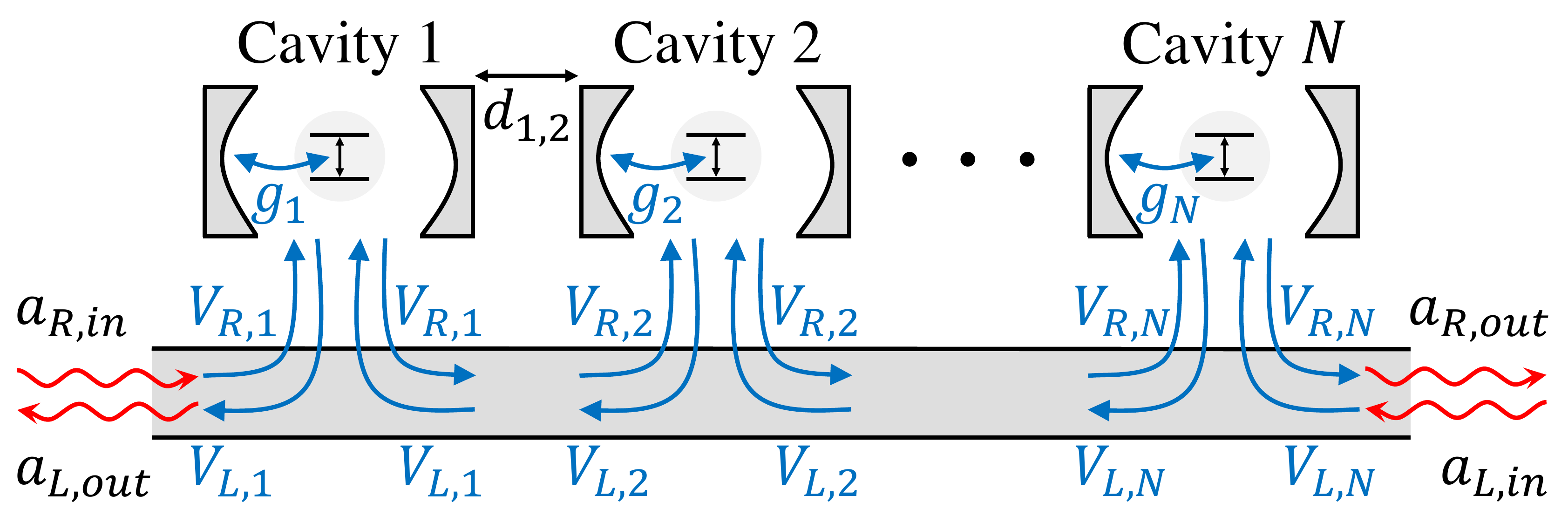}
    \caption{$N$ single-mode cavities coupled to a single-mode waveguide, with each cavity containing a two-level emitter. Photon propagation is indicated by red wavy arrows, and coupling is represented by blue arrows. Cavity $j$ couples to its emitter with coupling rate $g_j$, and to the right- and left-moving waveguide modes with coupling rates $V_{R,j}$ and $V_{L,j}$, respectively. The separation between two neighbouring cavities $(i,j)$ is $d_{i,j}$. The input and output modes are $a_{R,\text{in}}$, $a_{L,\text{in}}$ and $a_{R,\text{out}}$, $a_{L,\text{out}}$, respectively.}
    \label{fig:N_cavities_diagram}
\end{figure}

In this paper, we propose a single-photon switch with high efficiency and fidelity based on an array of single-mode cavities coupled to a common single-mode waveguide, with each cavity containing a two-level quantum emitter (Fig.~\ref{fig:N_cavities_diagram}). The switch is an optical router triggered by a classical signal --- the control of the propagation direction of a single input photon in the waveguide is realised by controlling the emitter-cavity coupling within each cavity. The waveguide allows for larger cavity separations compared to conventional CCAs with evanescent cavity-cavity coupling, while retaining strong inter-cavity interactions~\cite{Sato2012}. This makes individual cavity addressability more practical, enabling more precise control over photon propagation. For example, the cavity mode frequencies and the transition frequencies of the emitters can be tuned independently at each cavity site, e.g., via optical~\cite{Bose2011} or electrical~\cite{Nowak2014} Stark tuning, or through strain tuning~\cite{Grim2019, Luxmoore2012}. This tuning enables the emitter-cavity coupling to be controlled within each cavity, as required for the switch operation. Furthermore, the cavity-waveguide separations provide an additional degree of freedom that allows for independent control over inter-cavity coupling rates between any pair of cavities (e.g., by adjusting the separations electromechanically~\cite{Chew2010, Ohta2013}). The enhanced control in this system overcomes a major challenge in fabricating CCA-based single-photon routers with evanescent nearest-neighbour coupling, where identical cavities are required for good photon transmission but overcoming disorder is exceedingly difficult.


\section{theory}\label{sec:theory}

A schematic of the system is shown in Fig.~\ref{fig:N_cavities_diagram}. $N$ single-mode cavities are coupled to a common single-mode waveguide. Cavity $j$ has resonance frequency $\omega_{c,j}$ and contains a two-level quantum emitter with transition frequency $\omega_{e,j}$ and emitter-cavity coupling rate $g_j$. The cavity couples to the right- and left-moving waveguide modes with coupling rates $V_{R,j}$ and $V_{L,j}$, respectively. The separation between two neighbouring cavities $(i,j)$ is $d_{i,j}$. We can write the Hamiltonian for cavity $j$ as:
\begin{equation}
H_j = H_e + H_c + H_{\text{wg}} + H_{e\text{-}c} + H_{c\text{-wg}},
\label{eq:Hj}
\end{equation}
where ($\hbar = 1$)
\begin{equation}
H_e = \frac{1}{2}\omega_{e,j}\sigma_{z,j}
\end{equation}
is the free Hamiltonian for the two-level emitter in cavity $j$ (with Pauli operator $\smash{\sigma_{z,j} = \ketbra{e_j}{e_j} - \ketbra{g_j}{g_j}}$, where $\ket{g_j}$ is the ground state and $\ket{e_j}$ is the excited state of the emitter),
\begin{equation}
H_c = \omega_{c,j}^{\vphantom{\dag}}c_j^{\dag}c_j^{\vphantom{\dag}}
\end{equation}
is the free Hamiltonian for the cavity (with mode operators $c_j^{\vphantom{\dag}}$, $c_j^{\dag}$), and
\begin{equation}
H_{\text{wg}} = \int_0^{\infty} \omega(k) a_{R}^{\dag}(k) a_{R}^{\vphantom{\dag}}(k) dk + \int_{-\infty}^0 \omega(k) a_{L}^{\dag}(k) a_{L}^{\vphantom{\dag}}(k) dk
\end{equation}
is the free waveguide Hamiltonian, with $\smash{a_R^{\vphantom{\dag}}(k)}$, $\smash{a_R^{\dag}(k)}$, $\smash{a_L^{\vphantom{\dag}}(k)}$, $\smash{a_L^{\dag}(k)}$ being the $k$-space operators for the right- and left-moving waveguide modes, and $\omega(k)$ is the dispersion relation. In addition,
\begin{equation}
H_{e\text{-}c} = g_j^{\vphantom{*}} \sigma_j^+ c_j^{\vphantom{\dag}} + g_j^* \sigma_j^- c_j^{\dag}
\end{equation}
is the Jaynes-Cummings emitter-cavity interaction~\cite{Jaynes1963, Larson2021} (where $\smash{\sigma_j^+ = \ketbra{e_j}{g_j}}$ and $\smash{\sigma_j^- = \ketbra{g_j}{e_j}}$ are the raising and lowering operators for emitter $j$), and
\begin{align}
\begin{split}
H_{c\text{-wg}} = &\int_0^{\infty} \left[ \sqrt{\frac{V_{R,j}}{2\pi}} a_{R}^{\dag}(k) c_{j}^{\vphantom{\dag}} + \sqrt{\frac{V_{R,j}^*}{2\pi}} a_{R}^{\vphantom{\dag}}(k) c_{j}^{\dag} \right] dk\\[0.05in]
&+ \int_{-\infty}^0 \left[ \sqrt{\frac{V_{L,j}}{2\pi}} a_{L}^{\dag}(k) c_{j}^{\vphantom{\dag}} + \sqrt{\frac{V_{L,j}^*}{2\pi}} a_{L}^{\vphantom{\dag}}(k) c_{j}^{\dag} \right] dk
\end{split}
\end{align}
is the cavity-waveguide interaction, assumed to be independent of the photon wave number $k$ in the waveguide (Markov approximation~\cite{Gardiner1985}). There is no direct cavity-cavity interaction term (unlike in the Bose-Hubbard model, which is used to describe conventional CCAs~\cite{Hartmann2008}) because we consider sufficiently large separations $d_{i,j}$ where evanescent coupling between the cavities is exponentially suppressed.

The input and output modes of the waveguide are denoted by $a_{R,\text{in}}$, $a_{L,\text{in}}$ and $a_{R,\text{out}}$, $a_{L,\text{out}}$, respectively (see Fig.~\ref{fig:N_cavities_diagram}), and obey the linear transfer relation
\begin{equation}
\begin{pmatrix}
    a_{R,\text{out}} \\
    a_{L,\text{in}}
\end{pmatrix}
= T_{\text{tot}}
\begin{pmatrix}
    a_{R,\text{in}} \\
    a_{L,\text{out}}
\end{pmatrix},
\label{eq:transfer}
\end{equation}
where $T_{\text{tot}}$ is the total transfer matrix for the $N$-cavity system, which relates the input and output modes $a_{L,\text{in}}$, $a_{R,\text{out}}$ on the right side of cavity $N$ to the input and output modes $a_{R,\text{in}}$, $a_{L,\text{out}}$ on the left side of cavity $1$. Since the system in Fig.~\ref{fig:N_cavities_diagram} is an alternating sequence of waveguide-coupled cavities and regions of length $d_{i,j}$ where photons propagate freely in the waveguide, the total transfer matrix can be decomposed into a product of transfer matrices $T_j$ for the cavities and $\smash{T_{\text{wg}}^{(i,j)}}$ for the waveguide segments of length $d_{i,j}$ separating the cavities:
\begin{equation}
T_{\text{tot}} = T_N \cdots T_2 \hspace{0.025in} T_{\text{wg}}^{(1,2)} \hspace{0.025in} T_1.
\label{eq:Ttot}
\end{equation}

In Appendix~\ref{sec_app:A}, we transform the single-cavity Hamiltonian $H_j$ in Eq.~(\ref{eq:Hj}) from $k$-space to frequency-space using the linear dispersion approximation. We then use the resulting Hamiltonian and the input-output formalism~\cite{Gardiner1985} to derive the cavity transfer matrices $T_j$, by assuming a weak coherent input field in the waveguide (see Appendix~\ref{sec_app:B}). Here, we use the weak-excitation approximation to neglect multi-photon contributions in the coherent input~\cite{Rephaeli2012}, allowing us to consider single-photon scattering. This leads to the result
\begin{equation}
T_j = \frac{1}{\beta_j + \alpha_j^{(-)}}
\begin{pmatrix}
    \beta_j - \alpha_j^{(+)} & \zeta_j \\
    \zeta_j^* & \beta_j + \alpha_j^{(+)} \\
\end{pmatrix},
\label{eq:Tj}
\end{equation}
where $\smash{\alpha_j^{(\pm)} = \frac{i}{2}\left( |V_{R,j}| \pm |V_{L,j}| \right)}$, $\smash{\beta_j = \Delta_{c,j} - |g_j|^2/\Delta_{e,j}}$, and ${\zeta_j = -i\left( V_{R,j}^{\vphantom{*}} V_{L,j}^* \right)^{\frac{1}{2}}}$. Here, ${\Delta_{c,j} = \omega - \omega_{c,j}}$ and ${\Delta_{e,j} = \omega - \omega_{e,j}}$ are frequency detunings, where $\omega$ is the input photon frequency. Photon loss from the cavities and emitters into the environment can be included in this result using the substitutions ${\Delta_{c,j} \rightarrow \Delta_{c,j} + i\kappa_j/2}$ and ${\Delta_{e,j} \rightarrow \Delta_{e,j} + i\gamma_j/2}$, where $\kappa_j$ and $\gamma_j$ are the loss rates of cavity $j$ and emitter $j$, respectively (see Appendix~\ref{sec_app:C}).

The waveguide transfer matrices ${T_{\text{wg}}^{(i,j)}}$ describe phase shifts that the photons acquire when they propagate freely over the distances $d_{i,j}$ in the waveguide, so they have the simple form
\begin{equation}
T_{\text{wg}}^{(i,j)} = 
\begin{pmatrix}
    e^{-i \omega d_{i,j}/v_g} & 0 \\[0.05in]
    0 & e^{i \omega d_{i,j}/v_g} \\
\end{pmatrix},
\label{eq:Twg}
\end{equation}
where $v_g$ is the photon group velocity (this is also derived in Appendix~\ref{sec_app:B}).

Since the cavity transfer matrices $T_j$ and the waveguide transfer matrices $\smash{T_{\text{wg}}^{(i,j)}}$ are ${2 \times 2}$ matrices, after computing the product in Eq.~(\ref{eq:Ttot}) for some chosen number of cavities $N$, the total transfer matrix will also be a ${2 \times 2}$ matrix, with some matrix elements $M_{nm}$ that depend on the system parameters:
\begin{equation}
T_{\text{tot}} = 
\begin{pmatrix}
    M_{11} & M_{12} \\
    M_{21} & M_{22} \\
\end{pmatrix}.
\label{eq:Ttot_matrix}
\end{equation}
Substituting this matrix into the transfer relation in Eq.~(\ref{eq:transfer}) allows the output modes $a_{R,\text{out}}$, $a_{L,\text{out}}$ to be expressed in terms of the input modes $a_{R,\text{in}}$, $a_{L,\text{in}}$:
\begin{subequations}
\begin{equation}
a_{R,\text{out}} = \left( M_{11} - \frac{M_{12} M_{21}}{M_{22}} \right) a_{R,\text{in}} + \frac{M_{12}}{M_{22}} a_{L,\text{in}},
\end{equation}
\begin{equation}
a_{L,\text{out}} = -\frac{M_{21}}{M_{22}} a_{R,\text{in}} + \frac{1}{M_{22}} a_{L,\text{in}}.
\end{equation}
\end{subequations}
For right-moving input photons, the coefficient of $a_{R,\text{in}}$ in the expression for $a_{R,\text{out}}$ is the $N$-cavity transmission coefficient $t_N$ (this is the probability amplitude for the output photons also moving to the right), and the coefficient of $a_{R,\text{in}}$ in the expression for $a_{L,\text{out}}$ is the $N$-cavity reflection coefficient $r_N$ (this is the probability amplitude for the output photons being scattered to the left):
\begin{equation}
t_N = M_{11} - \frac{M_{12}M_{21}}{M_{22}}
\quad\text{and}\quad
r_N = - \frac{M_{21}}{M_{22}}.
\label{eq:tN_and_rN}
\end{equation}
These equations allow transmission and reflection spectra to be calculated for any number of waveguide-coupled cavities, once the elements $M_{nm}$ of the total transfer matrix $T_{\text{tot}}$ are computed using Eq.~(\ref{eq:Ttot}).


\section{Results}\label{sec:results}


\subsection{Transmission Spectra for Different $N$}\label{subsec:results_A}

\begin{figure}
    \includegraphics[width = \columnwidth]{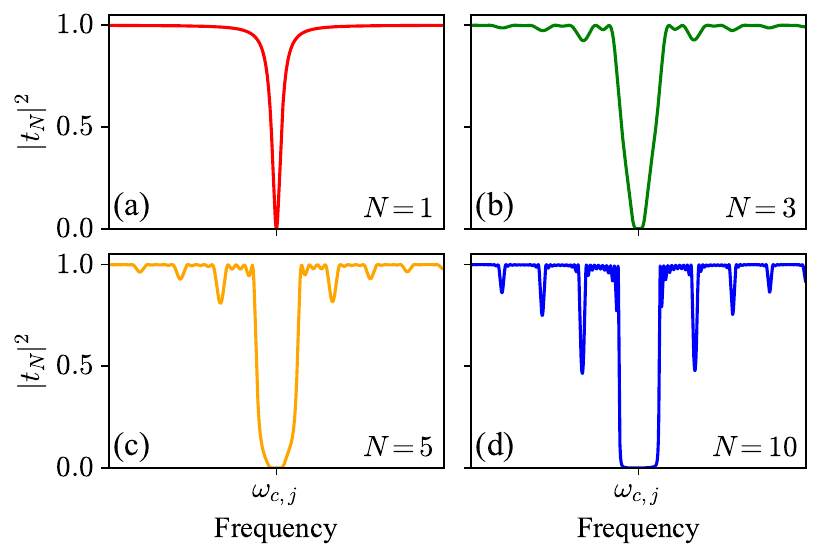}
    \caption{Transmission $|t_N|^2$ as a function of the input photon frequency for (a) ${N=1}$, (b) ${N=3}$, (c) ${N=5}$ and (d) ${N=10}$ identical, equally-spaced waveguide-coupled cavities in the weak emitter-cavity coupling regime (${g_j \ll \kappa_j, \gamma_j}$ for all $j$).}
    \label{fig:transmission_different_N}
\end{figure}

In Fig.~\ref{fig:transmission_different_N}, we show the transmission $|t_N|^2$ as a function of the input photon frequency for (a) ${N=1}$, (b) ${N=3}$, (c) ${N=5}$ and (d) ${N=10}$ identical, equally-spaced cavities (in Appendix~\ref{sec_app:D}, we derive analytical results for $t_N$ and $r_N$ for any number $N$ of identical, equally-spaced cavities). Here, all the cavities are in the weak emitter-cavity coupling regime (${g_j \ll \kappa_j, \gamma_j}$), so the effect of the emitters is negligible. For a single waveguide-coupled cavity [Fig.~\ref{fig:transmission_different_N}(a)], a transmission dip occurs at the cavity resonance frequency~\cite{Waks2006, Shen2009}. When the number of cavities is increased [Figs.~\ref{fig:transmission_different_N}(b)-(d)], the dip at the cavity resonance frequency $\omega_{c,j}$ is broadened compared to the ${N=1}$ case. This transmission dip broadening is consistent with previous observations in systems with multiple emitters or resonators coupled to a common waveguide~\cite{Tsoi2009, Yan2018_2, Ren2022, Berndsen2024}. Furthermore, for ${N>1}$ we see fringes that arise from interference in the waveguide, caused by reflections between the cavities. This interference leads to a periodic sequence of transmission dips on either side of the central dip, with a free spectral range ${\Delta_{\text{FSR}} = 1/2T_P = v_g/2d_{i,j}}$~\cite{Sato2012}, where ${T_P = d_{i,j}/v_g}$ is the photon propagation time between neighbouring cavities. We note that our system enables the free spectral range to be controlled post-fabrication, for example by decoupling every other cavity from the waveguide to increase the nearest-neighbour separations $d_{i,j}$ by a factor of two, which would reduce $\Delta_{\text{FSR}}$ by a factor of two. The interference in a waveguide-based CCA can also give rise to photon antibunching in the output of the waveguide~\cite{Lu2024}.


\subsection{Photon Switch Operation}\label{subsec:results_B}

\begin{figure}
    \includegraphics[width = \columnwidth]{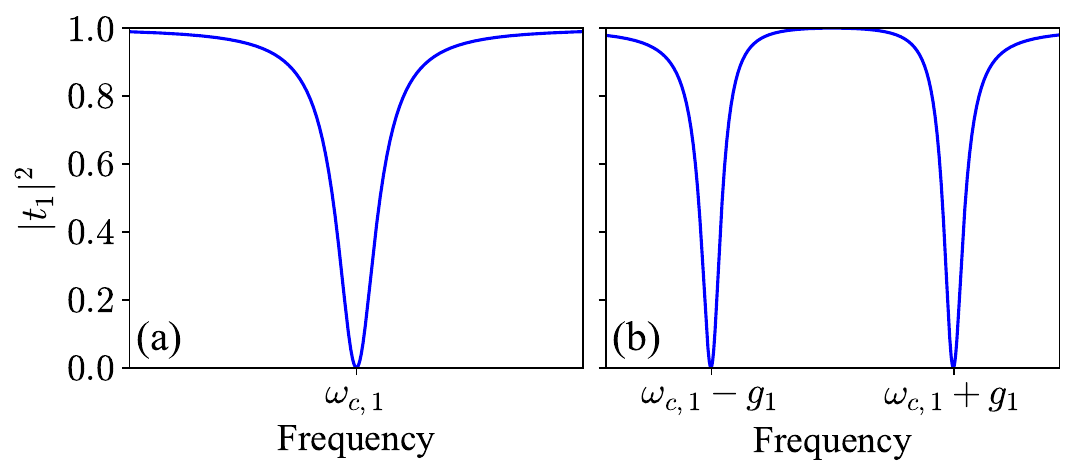}
    \caption{Transmission $|t_{1}|^2$ as a function of the input photon frequency for a single waveguide-coupled cavity. (a) and (b) correspond to the weak (${g_1 \ll \kappa_1, \gamma_1}$) and strong (${g_1 \gg \kappa_1, \gamma_1}$) emitter-cavity coupling regimes, respectively.}
    \label{fig:switch_1_cavity}
\end{figure}

We now discuss the basic working principle of our photon switch proposal. As shown in Fig.~\ref{fig:transmission_different_N}, when the cavities are in the weak coupling regime (${g_j \ll \kappa_j, \gamma_j}$), there is a dip in transmission centred at the cavity resonance frequencies, where input photons can be reflected. When we transition to the strong emitter-cavity coupling regime (${g_j \gg \kappa_j, \gamma_j}$), the coherent interaction between the emitters and the cavities induces vacuum Rabi splitting of the cavity modes~\cite{Khitrova2006}, which opens a transmission window that allows photons to propagate through the waveguide. We show this for a single waveguide-coupled cavity in Fig.~\ref{fig:switch_1_cavity}. In the strong coupling regime [Fig.~\ref{fig:switch_1_cavity}(b)], the transmission dip splits into two dips with a frequency separation of $2g_1$. Due to this splitting, the transmission at the cavity resonance frequency $\omega_{c,1}$ switches from being zero to approximately one when we switch from weak coupling to strong coupling~\cite{Waks2006, Shen2009}. This means that we can deterministically switch between reflection and transmission by controlling the emitter-cavity coupling rate $g_1$. Experimentally, this can be achieved by controlling the emitter-cavity detuning, for example using applied fields~\cite{Bose2011, Nowak2014}.

\begin{figure}
    \includegraphics[width = \columnwidth]{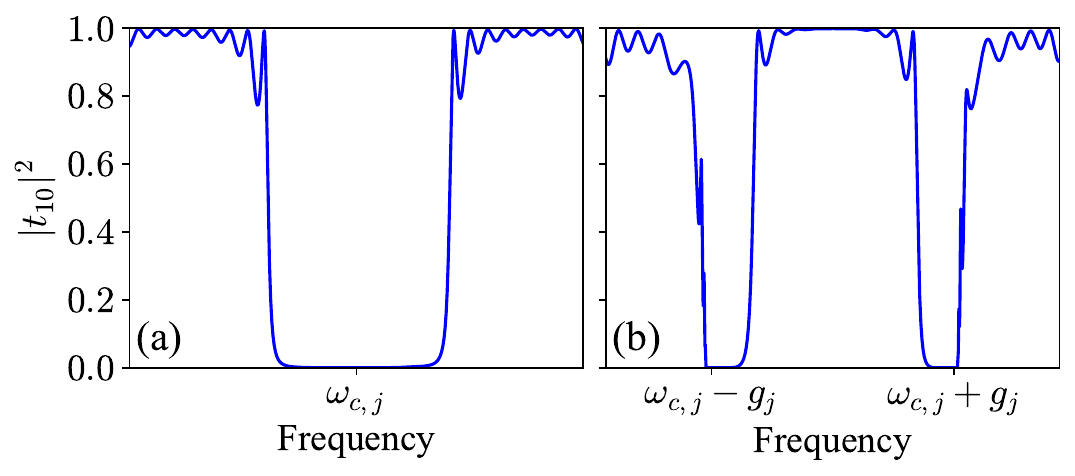}
    \caption{Transmission $|t_{10}|^2$ as a function of the input photon frequency for ten identical, equally-spaced waveguide-coupled cavities. (a) and (b) correspond to the weak (${g_j \ll \kappa_j, \gamma_j}$) and strong (${g_j \gg \kappa_j, \gamma_j}$) emitter-cavity coupling regimes, respectively.}
    \label{fig:switch_10_cavities}
\end{figure}

Fig.~\ref{fig:switch_1_cavity} illustrates how a single waveguide-coupled cavity can be used as a photon switch for input photons in the waveguide. In the weak coupling regime [Fig.~\ref{fig:switch_1_cavity}(a)], perfect reflection occurs only at the cavity resonance frequency, allowing only narrow wave packets centred at this frequency to be routed effectively. However, we can increase the switching bandwidth using the behaviour observed in Fig.~\ref{fig:transmission_different_N}, where increasing the number of cavities was shown to increase the width of the transmission dip in the weak coupling regime. In Fig.~\ref{fig:switch_10_cavities}, we show the operation of the switch for ${N=10}$ identical, equally-spaced cavities, for comparison with the ${N=1}$ case from Fig.~\ref{fig:switch_1_cavity}. Fig.~\ref{fig:switch_10_cavities}(a) corresponds to the weak coupling regime (${g_j \ll \kappa_j, \gamma_j}$ for all ${j \in \{1,\dotsc,10\}}$) --- this is the same as Fig.~\ref{fig:transmission_different_N}(d) but with only the central transmission dip shown. When all the emitter-cavity pairs are in the strong coupling regime [Fig.~\ref{fig:switch_10_cavities}(b), ${g_j \gg \kappa_j, \gamma_j}$ for all $j$], we observe Rabi splitting as in the ${N=1}$ case, which results in two dips with a frequency separation of $2g_j$ and a region of high transmission between them. Importantly, we can now switch between near-unity reflection and near-unity transmission over a range of frequencies centred at the cavity resonance frequency $\omega_{c,j}$, as opposed to switching only at one frequency. Using multiple waveguide-coupled cavities therefore increases the switching bandwidth, allowing photon wave packets with a wider frequency distribution to be routed in the desired direction in the waveguide.

The possibility of using larger cavity separations in our system allows us to overcome fabrication disorder with greater ease compared to CCA proposals with evanescent nearest-neighbour coupling. Nevertheless, it may not be possible to make all cavities and emitters identical in a given implementation (e.g., due to a limited tuning range). Hence, in Appendix~\ref{sec_app:E}, we analyse how the ideal ten-cavity spectra in Fig.~\ref{fig:switch_10_cavities} are affected by disorder. We find that the spectra are highly robust against variations in the quality ($Q$) factors of the cavities, even when there is a $25\%$ standard deviation in the $Q$ factors. The switching is also robust against sub-wavelength disorder in the cavity separations $d_{i,j}$, but the performance can degrade significantly with wavelength-scale disorder due to the interferometric nature of the switch. In addition, the transmission spectra cope well with disorder in the cavity resonance frequencies $\omega_{c,j}$ and emitter transition frequencies $\omega_{e,j}$, provided that the frequency distributions do not exceed the switching bandwidth. Furthermore, we find that, if the strong coupling regime cannot be reached in one of the cavities in the array (e.g., due to poor emitter positioning within the cavity), then such a cavity should be detuned away from the switching region to recover the desired transmission behaviour (effectively reducing the number of cavities in the switch by one). This is a general mitigation strategy we can employ in our system --- if there are cavities where strong coupling cannot be achieved due to fabrication imperfections, then such cavities should be detuned away from the switching bandwidth (or decoupled from the waveguide) for the operation of the switch to work as intended (for examples of cavity tuning mechanisms, see~\cite{Luxmoore2012, Chew2010, Ohta2013, Marki2006, Gil-Santos2017, Zhou2020}). We note that this would not be possible in a conventional CCA with evanescent nearest-neighbour coupling, where detuning one cavity away from the rest would inhibit photon transmission through the array (in our system, photon transport does not rely on all the cavities being in resonance). When we decouple a cavity from the waveguide, we introduce some disorder in the cavity separations, but we find that this does not have a negative impact on the transmission window. The impact would be more significant in the weak coupling regime (where the switching window depends more strongly on the cavity number), and hence in this regime it would be more favourable to have all the cavities coupled to the waveguide and tuned to the centre of the switching bandwidth.


\subsection{Switch Efficiencies and Fidelities}\label{subsec:results_C}

From Figs.~\ref{fig:switch_1_cavity} and \ref{fig:switch_10_cavities}, we see that the switch operates in reflection mode (`$r$') in the weak emitter-cavity coupling regime, and in transmission mode (`$t$') in the strong coupling regime. We quantify the performance of our proposed switch by calculating the efficiency $E_{\nu}$ and fidelity $F_{\nu}$ in both regimes using
\begin{subequations}
\begin{equation}
E_{\nu} = \left| \int_{-\infty}^{\infty} \left|\nu_N(\omega)\right|^2 \left|f(\omega)\right|^2 d\omega \right|^2,
\label{eq:E}
\end{equation}
\begin{equation}
F_{\nu} = \left| \int_{-\infty}^{\infty}\nu_N(\omega) \left|f(\omega)\right|^2 d\omega \right|^2,
\label{eq:F}
\end{equation}
\end{subequations}
where ${\nu \in \{r,t\}}$, and we consider a Gaussian input photon wave packet ${f(\omega)}$ (see Appendix~\ref{sec_app:F} for the wave packet and the derivations of the efficiency and fidelity expressions). The efficiency is the probability that the wave packet will leave in the desired direction in the waveguide, while the fidelity quantifies the similarity of the input and output wave packets. For an ideal switch, ${E_\nu = 1}$ (input wave packets always leave in the correct direction --- deterministic operation) and ${F_\nu = 1}$ (the wave packet shape is preserved --- faithful operation) in both reflection and transmission mode.

\begin{figure}
    \includegraphics[width = \columnwidth]{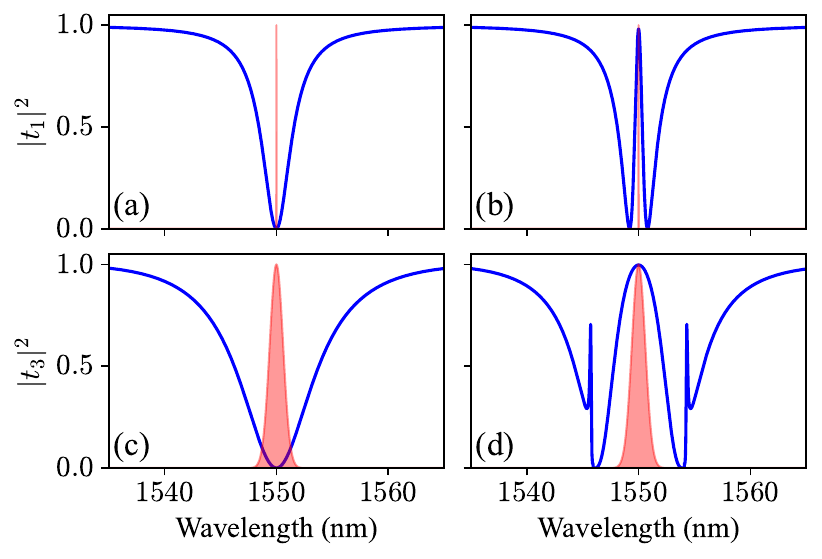}
    \caption{Examples of switching different wave packets. The wave packets correspond to the red shaded regions (rescaled to have unit height), and the blue curves show transmission as a function of the input photon wavelength. In (a) and (b), we consider a switch with a single waveguide-coupled cavity (${N=1}$), and the input Gaussian wave packet has a central wavelength of $1550$~nm and a full width at half-maximum ${\sigma_{\lambda} = 0.01}$~nm. In (c) and (d), we consider a switch with three cavities (${N=3}$), and the input wave packet has a central wavelength of $1550$~nm and a full width at half-maximum ${\sigma_{\lambda} = 1}$~nm.}
    \label{fig:transmission_switch}
\end{figure}

Fig.~\ref{fig:transmission_switch} shows examples of switching different wave packets. The parameters we use for the efficiency and fidelity calculations correspond to photon wavelengths in the telecom C-band, which are relevant for quantum networks with applications in quantum communications (we note however that the parameters can be rescaled, and the results generalised to other photon wavelengths). In particular, the cavity mode wavelengths are ${\lambda_{c,j} = 1550}$~nm (${\omega_{c,j} = 2\pi c/\lambda_{c,j}}$, where $c$ is the vacuum speed of light), and ${Q_{c,j} = 500}$ are the $Q$ factors of the cavities when coupled to the waveguide, giving the cavity-waveguide coupling rates ${V_{R,j} = V_{L,j} = \omega_{c,j}/2Q_{c,j}}$ (here, the factor of ${1/2}$ arises because we consider equal coupling to the right- and left-moving waveguide modes). The cavity loss rates into non-guided modes are ${\kappa_j = \omega_{c,j}/Q_{u,j}}$, where ${Q_{u,j} = 5\times10^4}$ are the intrinsic $Q$ factors of the cavities when not coupled to the waveguide (${\kappa_j/2\pi \approx 4}$~GHz), and ${\gamma_j/2\pi = 1}$~GHz are the emitter loss rates (typical for semiconductor quantum dot systems~\cite{Englund2012}).

In Figs.~\ref{fig:transmission_switch}(a) and (b), we use a single waveguide-coupled cavity to switch an input Gaussian wave packet centred at $1550$~nm (i.e., at the centre of the switching region), with a full width at half-maximum (FWHM) ${\sigma_{\lambda} = 0.01}$~nm. This wave packet is representative of photon emission from a quantum dot~\cite{Phillips2024}. Fig.~\ref{fig:transmission_switch}(a) corresponds to the weak coupling regime, where ${g_1/2\pi = 100}$~MHz (${g_1 \ll \kappa_1, \gamma_1}$), and the wave packet is reflected with an efficiency ${E_r = 96.1\%}$ and a fidelity ${F_r = 98.0\%}$. In Fig.~\ref{fig:transmission_switch}(b), we increase the emitter-cavity coupling rate to ${g_1/2\pi = 100}$~GHz, with the emitter being on resonance with the cavity (i.e., ${\lambda_{e,1} = \lambda_{c,1} = 1550}$~nm). This results in a transition to the strong coupling regime (${g_1 \gg \kappa_1, \gamma_1}$), leading to a Rabi splitting of ${2g_1/2\pi = 200}$~GHz (${\approx 1.6}$~nm, or about $800$~$\upmu$eV). This splitting allows the wave packet to be transmitted through the waveguide with an efficiency ${E_t = 96.2\%}$ and a fidelity ${F_t = 98.1\%}$. We note that $E_t$ and $F_t$ approach $100\%$ when the emitter-cavity coupling strength $g_1$ is increased, but we use a coupling strength that is of the order of what has been achieved to date with quantum dots in semiconductor nanocavities~\cite{Ota2018}. We also note that the same efficiencies and fidelities can be attained in transmission with ${g_1/2\pi = 50}$~GHz and ${Q_{c,1} = 2000}$, i.e., by increasing the coupled-cavity $Q$ by a factor of four, we can achieve the same switching performance with half the emitter-cavity coupling rate. In practice, we can modulate the coupled-$Q$ factor in situ by adjusting the cavity-waveguide separation~\cite{Ohta2013}.

In the situation considered in Figs.~\ref{fig:transmission_switch}(a) and (b), the width of the wave packet is significantly smaller than the cavity mode linewidth [compare the wave packet width with the width of the transmission dip in Fig.~\ref{fig:transmission_switch}(a)]. Hence, a single waveguide-coupled cavity can route the wave packet with near-unity efficiency and fidelity, without the need for introducing more cavities to increase the width of the transmission dip. In Figs.~\ref{fig:transmission_switch}(c) and (d), we consider switching a Gaussian wave packet centred at $1550$~nm (again at the centre of the switching region), with a FWHM ${\sigma_{\lambda} = 1}$~nm (this wave packet corresponds to a few-picosecond pulse). Since this wave packet has a much larger spectral width than the wave packet considered in Figs.~\ref{fig:transmission_switch}(a) and (b), comparable to the cavity mode linewidth, we need to increase the switching bandwidth to maintain high routing efficiencies and fidelities. This can be achieved by coupling more cavities to the waveguide to increase the width of the reflection window in the weak coupling regime (see Fig.~\ref{fig:transmission_different_N}), and by using larger emitter-cavity coupling rates $g_j$ to increase the width of the transmission window in the strong coupling regime. Hence, in Figs.~\ref{fig:transmission_switch}(c) and (d), we use an array of three waveguide-coupled cavities to route the wider wave packet. Here, the nearest-neighbour cavity separations are ${d_{1,2} = d_{2,3} = 4.65}$~$\upmu$m (comparable to previous experiments involving waveguide-coupled nanocavities~\cite{Sato2012}), and the photon group velocity in the waveguide is ${v_g = 0.3c}$ (corresponding to the group index ${n_g = c/v_g \approx 3}$). Fig.~\ref{fig:transmission_switch}(c) corresponds to the weak coupling regime, where ${g_j/2\pi = 100}$~MHz for all ${j \in \{1,2,3\}}$. Comparing this with Fig.~\ref{fig:transmission_switch}(a), we see the broadening of the transmission dip due to using a larger number of cavities, resulting in a reflection efficiency ${E_r = 96.4\%}$ and a reflection fidelity ${F_r = 97.7\%}$ for the wider wave packet (if only one cavity with the same parameters was used to reflect this wave packet, we would obtain ${E_r = 84.7\%}$ and ${F_r = 86.4\%}$). Fig.~\ref{fig:transmission_switch}(d) corresponds to the strong coupling regime, where ${g_j/2\pi = 500}$~GHz for all $j$, and all the emitters are on resonance with the cavities (${\lambda_{e,j} = \lambda_{c,j} = 1550}$~nm). With these parameters, we obtain a transmission efficiency ${E_t = 96.4\%}$ and a transmission fidelity ${F_t = 97.4\%}$. As in the single-cavity case, we can increase the coupled-$Q$ factors $Q_{c,j}$ (e.g., by increasing the cavity-waveguide separations) to achieve similar transmission efficiencies and fidelities with lower coupling rates. For example, with ${g_j/2\pi = 250}$~GHz and ${Q_{c,j} = 2000}$, we obtain ${E_t = 95.5\%}$ and ${F_t = 96.6\%}$. In general, the switching bandwidth needs to be increased compared to the ${N=1}$ case if the wave packet width is comparable to or greater than the cavity mode linewidth, so it is necessary to use multiple cavities and larger emitter-cavity coupling rates to maintain near-unity efficiencies and fidelities. The coupling rates used here for the strong coupling regime correspond to Rabi splittings of several meV, and have been exceeded experimentally in systems such as quantum dots in plasmonic nanogap cavities~\cite{Santhosh2016} and excitons in cavities constructed from organic molecules~\cite{Tang2024}. Similar coupling rates to the values used here can also be reached by placing a two-level emitter inside a semiconductor nanogap cavity~\cite{Uemoto2014}, where loss rates are less significant than in plasmonic systems.

\begin{figure}
    \includegraphics[width = \columnwidth]{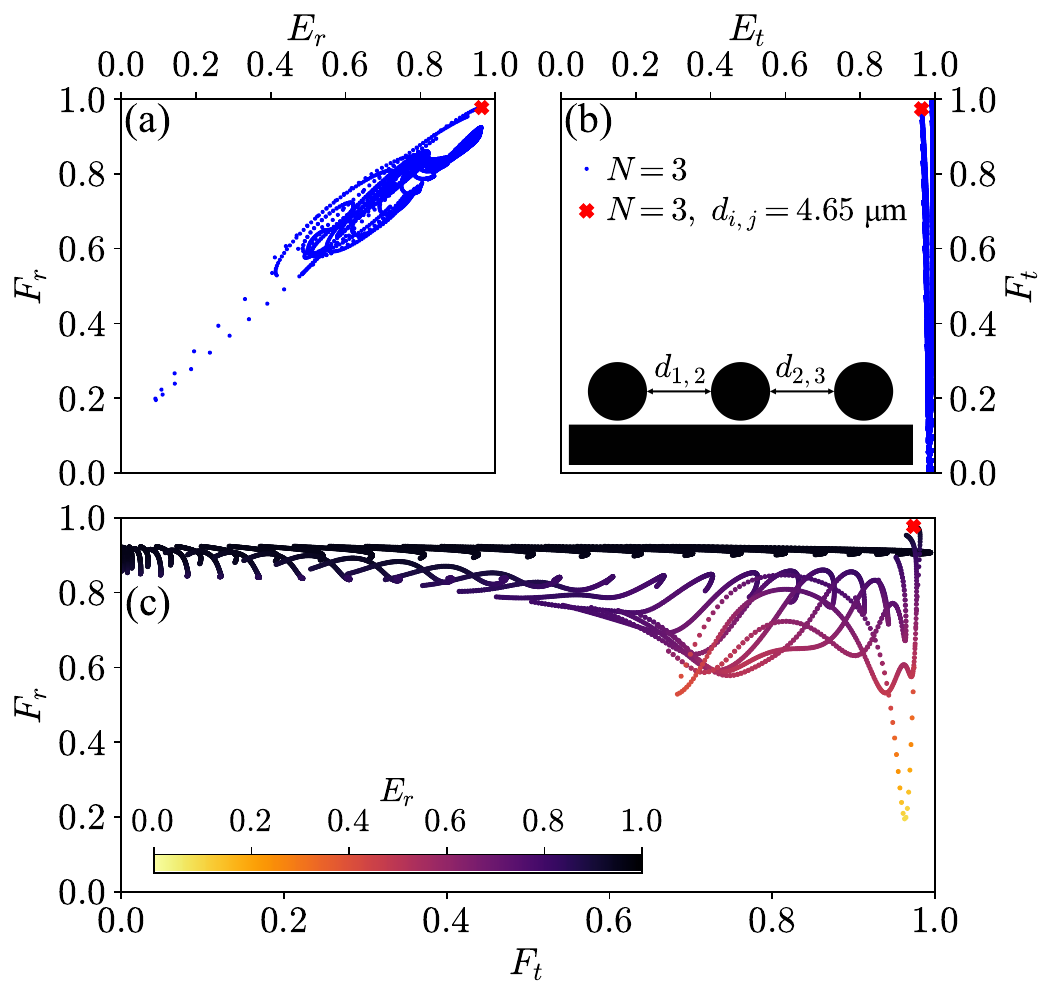}
    \caption{Switch efficiencies and fidelities for ${N=3}$ identical, equally-spaced cavities [as shown schematically in the inset in (b)]. The input Gaussian wave packet has a FWHM ${\sigma_{\lambda} = 1}$~nm, as in Figs.~\ref{fig:transmission_switch}(c) and (d). (a) Reflection efficiency $E_r$ and reflection fidelity $F_r$ calculated for cavity separations in the range ${d_{1,2} = d_{2,3} = 1\text{ - }100}$~$\upmu$m, with $0.01$~$\upmu$m increments (blue points). This data corresponds to the weak emitter-cavity coupling regime (${g_j/2\pi = 100}$~MHz), where the switch is in reflection mode. (b) Transmission efficiency $E_t$ and transmission fidelity $F_t$ in the strong coupling regime (${g_j/2\pi = 500}$~GHz), where the switch is operating in transmission mode. The same set of cavity separations is used to calculate the efficiencies and fidelities as in (a). (c) $F_r$ shown against $F_t$, using the same data as in (a) and (b). The colour gradient represents the corresponding values of $E_r$. In (a), (b), and (c), the red cross indicates the results for ${d_{1,2} = d_{2,3} = 4.65}$~$\upmu$m, which is the configuration considered in Figs.~\ref{fig:transmission_switch}(c) and (d).}
    \label{fig:fidelity_efficiency_vs_d}
\end{figure}

When multiple cavities are used, the performance of the switch depends greatly on the choice of the cavity separations $d_{i,j}$ due to its interferometric nature. In Fig.~\ref{fig:fidelity_efficiency_vs_d}, we show efficiencies and fidelities for different choices of $d_{i,j}$ in the three-cavity switch considered in Figs.~\ref{fig:transmission_switch}(c) and (d). In Figs.~\ref{fig:fidelity_efficiency_vs_d}(a) and (b), each blue point corresponds to a particular nearest-neighbour separation, which we vary in the range ${d_{i,j} = 1\text{ - }100}$~$\upmu$m in steps of $0.01$~$\upmu$m, keeping $d_{1,2}$ and $d_{2,3}$ equal [as shown schematically in the inset in Fig.~\ref{fig:fidelity_efficiency_vs_d}(b)]. The red cross indicates the situation considered in Figs.~\ref{fig:transmission_switch}(c) and (d), where ${d_{1,2} = d_{2,3} = 4.65}$~$\upmu$m. This is one of the best configurations for the three-cavity switch, but we see that there are separations where the efficiencies and fidelities are significantly reduced, and can be smaller than for a single waveguide-coupled cavity. This is because the interference in the waveguide can distort the transmission spectrum, including the switching bandwidth. This emphasises the importance of choosing the appropriate cavity separations to obtain the optimal performance. In Fig.~\ref{fig:fidelity_efficiency_vs_d}(c), we show the fidelities $F_r$ and $F_t$ using the data from Figs.~\ref{fig:fidelity_efficiency_vs_d}(a) and (b), where the colour gradient indicates the associated reflection efficiencies $E_r$ [here we do not show the transmission efficiencies $E_t$ as they are approximately equal for all separations, see Fig.~\ref{fig:fidelity_efficiency_vs_d}(b)]. The red cross again corresponds to the case where ${d_{1,2} = d_{2,3} = 4.65}$~$\upmu$m. We see that high fidelities in transmission and reflection can be achieved simultaneously in multiple configurations, not just the case indicated by the red cross [top-right corner in Fig.~\ref{fig:fidelity_efficiency_vs_d}(c)]. These high-fidelity points also correspond to high efficiencies, meaning that the switch is highly efficient and it preserves the input wave packet with near-unity fidelity in both reflection mode and transmission mode.

\begin{figure}
    \includegraphics[width = \columnwidth]{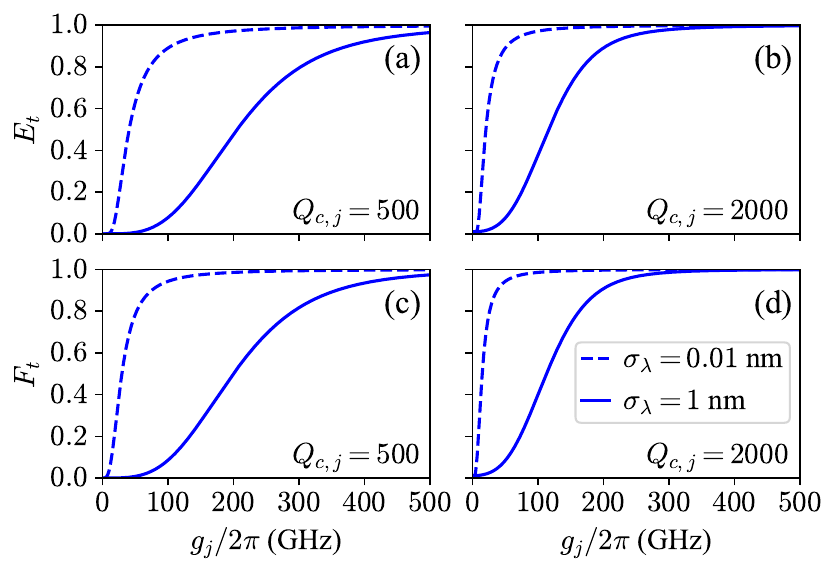}
    \caption{Transmission efficiency $E_t$ and fidelity $F_t$ as a function of the emitter-cavity coupling rates $g_j$ for a three-cavity switch. In (a) and (c) the coupled-$Q$ factors are ${Q_{c,j} = 500}$, while in (b) and (d) we use ${Q_{c,j} = 2000}$. The dashed curves correspond to an input wave packet that has a FWHM of $0.01$~nm [as in Figs.~\ref{fig:transmission_switch}(a) and (b)]. The solid curves correspond to an input wave packet that has a FWHM of $1$~nm [as in Figs.~\ref{fig:transmission_switch}(c) and (d)].}
    \label{fig:fidelity_efficiency_vs_g}
\end{figure}

In transmission mode, the larger the emitter-cavity coupling rates $g_j$, the wider the switching bandwidth, allowing wider wave packets to be transmitted through the waveguide with high efficiency and fidelity. In Fig.~\ref{fig:fidelity_efficiency_vs_g}, we show the transmission efficiency $E_t$ and fidelity $F_t$ as a function of the emitter-cavity coupling rates $g_j$ for the three-cavity switch with nearest-neighbour separations ${d_{1,2} = d_{2,3} = 4.65}$~$\upmu$m. The dashed curves correspond to an input photon wave packet that has a FWHM ${\sigma_{\lambda} = 0.01}$~nm [same wave packet as in Figs.~\ref{fig:transmission_switch}(a) and (b)], while the solid curves correspond to the wider wave packet from Figs.~\ref{fig:transmission_switch}(c) and (d), where the FWHM is $1$~nm. As expected, optimal efficiencies and fidelities can be achieved with lower coupling rates for the narrower wave packet, since a narrower transmission bandwidth (smaller Rabi splitting) is required. We also compare the case where the coupled-$Q$ factors are ${Q_{c,j} = 500}$ [Figs.~\ref{fig:fidelity_efficiency_vs_g}(a) and (c)] to the case where ${Q_{c,j} = 2000}$ [Figs.~\ref{fig:fidelity_efficiency_vs_g}(b) and (d)], showing that higher $Q$ factors allow higher transmission efficiencies and fidelities to be achieved with lower coupling rates for both wave packets.

\begin{figure}
    \includegraphics[width = \columnwidth]{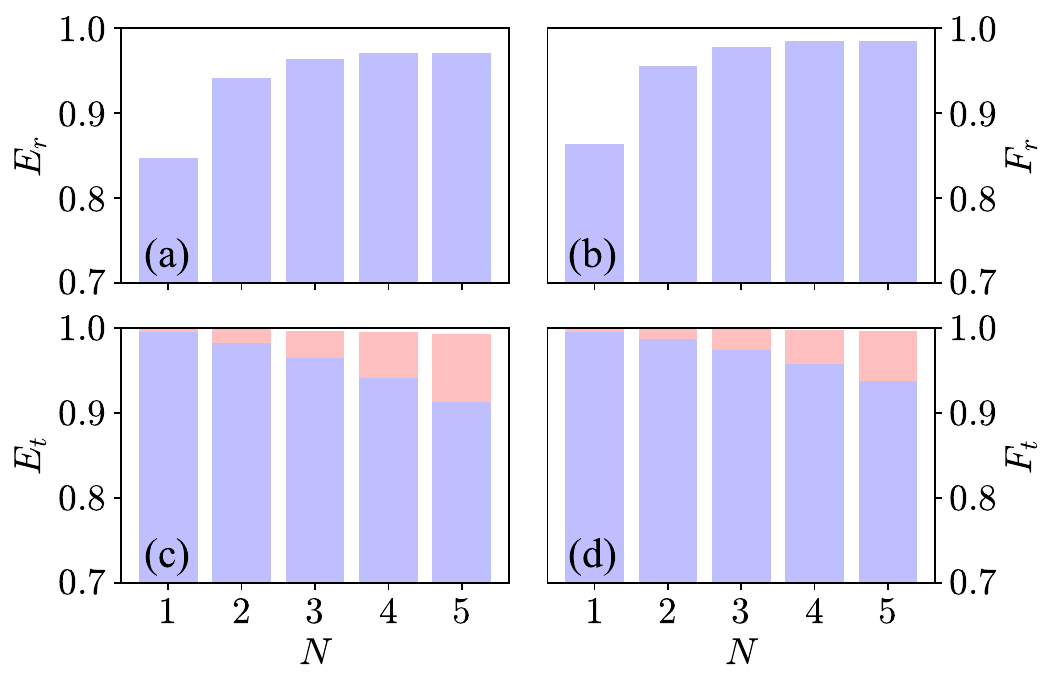}
    \caption{(a) Reflection efficiency $E_r$, (b) reflection fidelity $F_r$, (c) transmission efficiency $E_t$, and (d) transmission fidelity $F_t$ as a function of the number of cavities $N$. The input Gaussian wave packet has a FWHM of $1$~nm. (a) and (b) correspond to a switch in reflection mode (weak coupling regime, ${g_j/2\pi = 100}$~MHz), and (c) and (d) correspond to a switch in transmission mode (strong coupling regime, ${g_j/2\pi = 500}$~GHz), where ${Q_{c,j} = 500}$ for the blue bars and ${Q_{c,j} = 2000}$ for the red bars.}
    \label{fig:fidelity_efficiency_vs_N}
\end{figure}

In reflection mode, the switching bandwidth depends on the number $N$ of waveguide-coupled cavities used in the switch (as shown in Fig.~\ref{fig:transmission_different_N}). For wave packets wider than the cavity mode linewidth, it is beneficial to use multiple cavities, as this increases the reflection bandwidth in the weak coupling regime compared to the single-cavity case. In Fig.~\ref{fig:fidelity_efficiency_vs_N}, we show how the switch efficiencies and fidelities are affected when we change the number of cavities. In particular, we vary $N$ from $1$ to $5$, and we choose the cavity separations to be ${d_{i,j} = 4.65}$~$\upmu$m in all cases, as we found this to be one of the best configurations for the three-cavity switch (see Fig.~\ref{fig:fidelity_efficiency_vs_d}). We also consider the input wave packet with a FWHM of $1$~nm, as in the previous results for the three-cavity switch in Figs.~\ref{fig:transmission_switch}-\ref{fig:fidelity_efficiency_vs_g}. Figs.~\ref{fig:fidelity_efficiency_vs_N}(a) and (b) correspond to the weak coupling regime (${g_j/2\pi = 100}$~MHz for all $j$), where we see a noticeable improvement in the reflection efficiency $E_r$ and the reflection fidelity $F_r$ when we increase the number of cavities from ${N=1}$ to ${N=3}$. Beyond ${N=3}$, the improvement in $E_r$ and $F_r$ is very small, as increasing the width of the reflection window further does not provide a significant benefit for routing this particular wave packet. Figs.~\ref{fig:fidelity_efficiency_vs_N}(c) and (d) correspond to the strong coupling regime (${g_j/2\pi = 500}$~GHz), where ${Q_{c,j} = 500}$ for the blue bars and ${Q_{c,j} = 2000}$ for the red bars. We see that the transmission efficiency $E_t$ and the transmission fidelity $F_t$ decrease slightly when the number of cavities is increased, which is likely caused by the distortion of the transmission bandwidth due to interference [e.g., compare Fig.~\ref{fig:switch_1_cavity}(b) with Fig.~\ref{fig:switch_10_cavities}(b)]. This reduction in performance in transmission mode can be compensated by increasing the coupled-cavity $Q$ factors, as shown by the red bars (alternatively, larger emitter-cavity coupling rates can be used). In particular, we highlight that increasing the $Q$ factors from ${Q_{c,j} = 500}$ to ${Q_{c,j} = 2000}$ increases the transmission efficiency and fidelity of the switch with ${N=3}$ cavities from ${E_t = 96.4\%}$, ${F_t = 97.4\%}$ to ${E_t = 99.7\%}$, ${F_t = 99.8\%}$ (a similar increase could be achieved by keeping the $Q$ factors at ${Q_{c,j} = 500}$ and instead increasing the coupling rates to ${g_j/2\pi = 1}$~THz). We note that a higher $Q$ corresponds to a narrower cavity linewidth, and hence a narrower reflection bandwidth. Therefore, it is more favourable to use lower $Q$ factors for reflection mode and to increase the $Q$ factors when switching to transmission mode via in situ control rather than using high $Q$ factors in both regimes.


\subsection{Switching Speed}\label{subsec:results_D}

We now briefly consider how the width of the input photon wave packet affects the repetition rate of our proposed switch, as well as other mechanisms that can affect the switching speed. The spectral width of the wave packet will determine its time duration, and we require this duration to exceed the total round-trip time within the switch in order to observe the interference for ${N>1}$. Consider the three-cavity switch with nearest-neighbour cavity separations ${d_{1,2} = d_{2,3} = 4.65}$~$\upmu$m, and an input Gaussian wave packet with a FWHM of $1$~nm. The width ${\sigma_{\lambda} = 1}$~nm corresponds to a time duration of approximately $8$~ps for a pulse centred at $1550$~nm, while the photon round-trip time in the three-cavity switch is ${T = L/v_g = 18.6\text{ }\upmu\text{m}/0.3c \approx 0.2}$~ps, where ${L = 4.65\text{ }\upmu\text{m} \times 4 = 18.6~\upmu}$m is the total round-trip distance. When the time duration of the wave packet exceeds the round-trip time (as is the case here), it is the wave packet duration that limits the repetition rate of the switch. The time it takes to tune the emitter-cavity interactions between the weak and strong coupling regimes also determines how quickly the switch can be operated. The fastest tuning mechanisms include using an electric field to Stark shift the emitters on and off resonance with the cavities (this can enable switching at a rate of $150$~MHz~\cite{Faraon2010}), as well as optical shifting induced by a laser pump, which can enable switching on picosecond timescales~\cite{Bose2012, Volz2012, Fushman2007}. If the cavity $Q$ factors are to be modulated in order to improve the performance of the switch (instead of using larger emitter-cavity coupling rates), the speed with which the $Q$ factors can be controlled also needs to be considered.


\section{Conclusion}\label{sec:conclusion}

In conclusion, we have demonstrated theoretically that waveguide-coupled cavities with embedded quantum emitters can act as a highly efficient, high-fidelity single-photon switch. The switch reflects photons in the weak emitter-cavity coupling regime, and transmits photons in the strong coupling regime due to Rabi splitting. We find that a single waveguide-coupled cavity can reflect an input wave packet with a FWHM of $0.01$~nm (e.g., a photon emitted from a quantum dot single-photon source) with an efficiency ${E_r = 96.1\%}$ and a fidelity ${F_r = 98.0\%}$, or transmit the wave packet with an efficiency ${E_t = 96.2\%}$ and a fidelity ${F_t = 98.1\%}$. These values are achievable with parameters based on quantum dots in semiconductor nanostructures, for example a quantum dot embedded in a photonic crystal nanocavity. When the spectral width of the input wave packet is comparable to or greater than the cavity mode linewidth, the switching bandwidth needs to be increased to maintain high switching efficiencies and fidelities. This can be achieved by using multiple waveguide-coupled cavities to increase the reflection bandwidth in the weak coupling regime, and using larger emitter-cavity coupling rates to increase the transmission bandwidth in the strong coupling regime. For example, we find that an array of three waveguide-coupled cavities can reflect a wave packet with a FWHM of $1$~nm (corresponding to a few-picosecond pulse) with an efficiency ${E_r = 96.4\%}$ and a fidelity ${F_r = 97.7\%}$, and it can transmit the wave packet with an efficiency ${E_t = 99.7\%}$ and a fidelity ${F_t = 99.8\%}$. The switching between weak and strong coupling can be realised by controlling the emitter-cavity detuning within each cavity. Since the waveguide mediates inter-cavity coupling, the cavity separations can be significantly larger than the photon wavelength, allowing for independent control of emitter and cavity properties at each cavity site (e.g., with applied electric or optical fields). Our work shows that waveguide-coupled cavities with embedded emitters are a promising platform for the realisation of high-performance single-photon switches that preserve the input photon state with high fidelity, which is an essential requirement for photonic quantum technologies.


\section*{Acknowledgements}

The authors thank Elena Callus, Maxine M. McCarthy, Subhrajit Modak, and Nicholas J. Martin for helpful discussions. This work was supported by the Engineering and Physical Sciences Research Council [grant numbers EP/W524360/1, EP/V026496/1, EP/V021303/1, EP/M013472/1].


\appendix


\section{Hamiltonian Transformation}\label{sec_app:A}

\begin{figure}[b!]
    \includegraphics[width = 0.6\columnwidth]{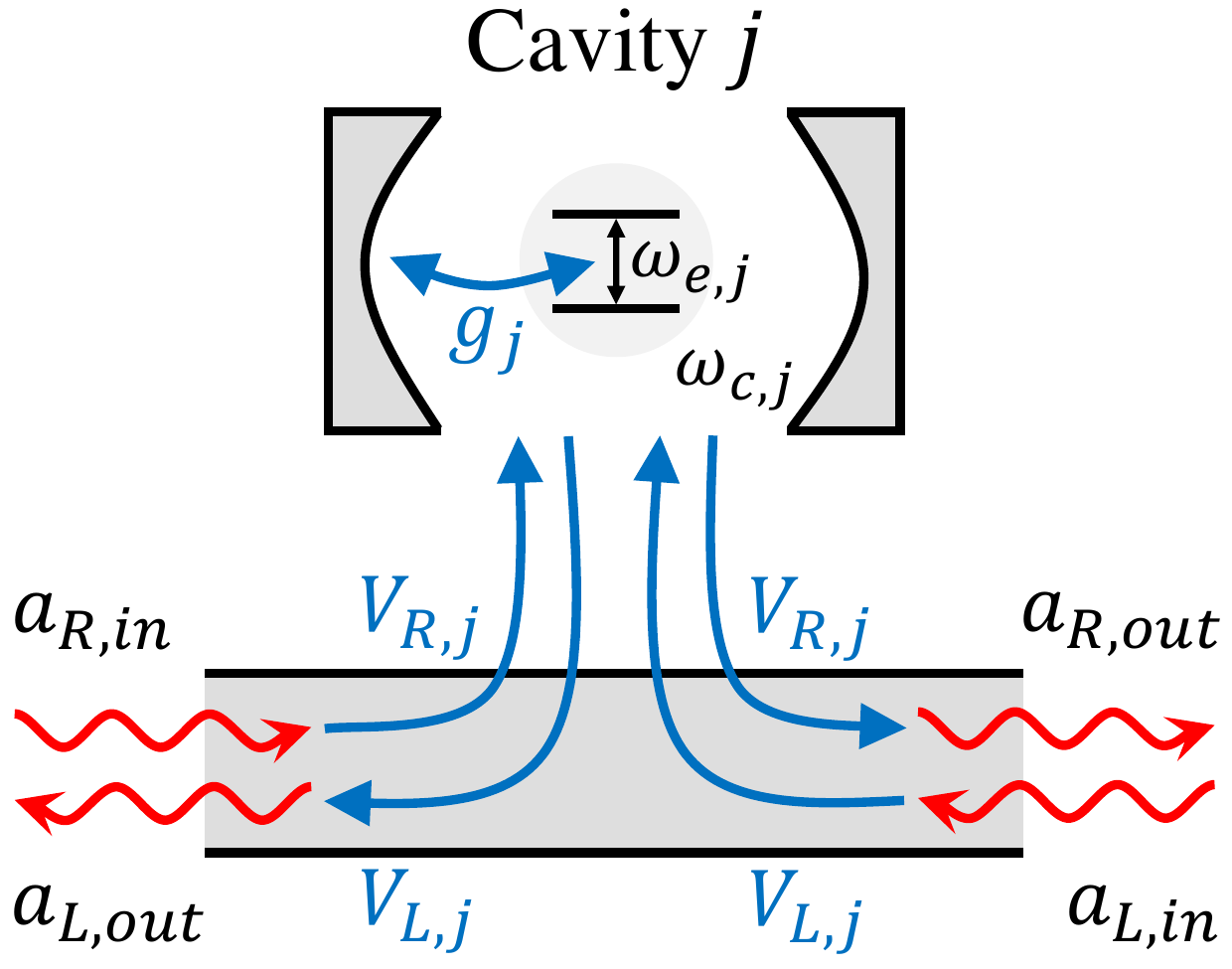}
    \caption{Diagram of a single waveguide-coupled cavity with the Hamiltonian $H_j$ given in Eq.~(\ref{eq:Hj}). The index $j$ labels the cavity.}
    \label{fig_app:single_cavity}
\end{figure}

\begin{figure}[b!]
    \includegraphics[width = 0.65\columnwidth]{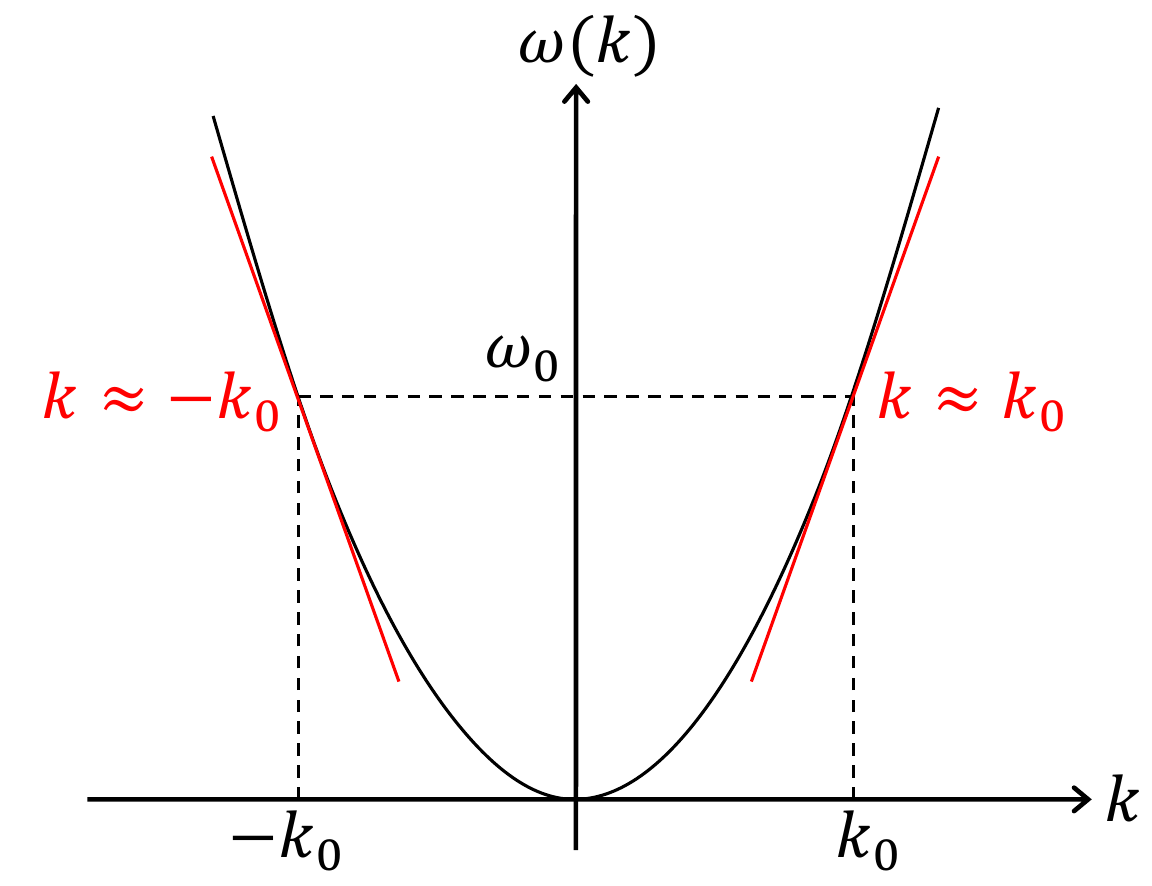}
    \caption{Graphical illustration of the linear dispersion approximation. The waveguide dispersion relation $\omega(k)$ (black curve) is approximated as being linear (red lines) near the wave numbers ${k = \pm k_0}$ corresponding to some frequency $\omega_0$.}
    \label{fig_app:linear_dispersion_approximation}
\end{figure}

In this appendix, we transform the single-cavity Hamiltonian $H_j$ in Eq.~(\ref{eq:Hj}) from $k$-space to frequency-space using the linear dispersion approximation, which is the first step in deriving the cavity transfer matrices $T_j$ using the input-output formalism. This Hamiltonian corresponds to the subsystem shown in Fig.~\ref{fig_app:single_cavity} (i.e., a single waveguide-coupled cavity from the full $N$-cavity system shown in Fig.~\ref{fig:N_cavities_diagram}). Within the linear dispersion approximation, we expand the waveguide dispersion relation $\omega(k)$ as a Taylor series around the wave numbers ${\pm k_0}$ corresponding to some frequency ${\omega_0 = \omega(k_0) = \omega(-k_0)}$, and we keep terms that are at most linear in $k$. We therefore have
\begin{equation}
\omega(k) \approx \left\{ \begin{array}{lr}
\omega_0 - v_g(k + k_0),\hspace{0.05in} k \approx -k_0, \\[0.05in]
\omega_0 + v_g(k - k_0),\hspace{0.05in}  k \approx k_0, \end{array}\right.
\end{equation}
where ${v_g = \frac{d\omega}{dk}\bigr|_{k_0}}$ is the photon group velocity in the waveguide. This linearisation is illustrated graphically in Fig.~\ref{fig_app:linear_dispersion_approximation}. For the Hamiltonian $H_j$, this approximation implies that ${\omega(k) \approx \omega_0 - v_g(k + k_0)}$ in the integrals over negative $k$, and ${\omega(k) \approx \omega_0 + v_g(k - k_0)}$ in the integrals over positive $k$, which in general is valid if only photons with wave numbers close to ${\pm k_0}$ are considered. However, since the linearisation point $\omega_0$ can be chosen freely and in our work we consider wave packets with a FWHM much smaller than the central frequency/wavelength, this is a well-justified approximation. With the above approximation for $\omega(k)$, the single-cavity Hamiltonian $H_j$ is given by the expression in Eq.~(\ref{eq_app:Hj_approx1}).
\begin{widetext}
\vspace{-0.2in}
\begin{align}
\begin{split}
H_j^{\vphantom{\dag}} =&\; \frac{1}{2}\omega_{e,j}^{\vphantom{\dag}}\sigma_{z,j}^{\vphantom{\dag}} + \omega_{c,j}^{\vphantom{\dag}}c_j^{\dag}c_j^{\vphantom{\dag}} + \int_0^{\infty}\Bigl[\omega_0 + v_g(k - k_0)\Bigr]a_R^\dag(k)a_R^{\vphantom{\dag}}(k)dk + \int_{-\infty}^0\Bigl[\omega_0 - v_g(k + k_0)\Bigr]a_L^\dag(k)a_L^{\vphantom{\dag}}(k)dk + g_j^{\vphantom{\dag}}\sigma_j^+c_j^{\vphantom{\dag}}\\[0.05in]
&+ g_j^{*\vphantom{\dag}}\sigma_j^-c_j^{\dag} + \int_0^{\infty}\left[ \sqrt{\frac{V_{R,j}^{\vphantom{*}}}{2\pi}}a_R^{\dag}(k)c_j^{\vphantom{\dag}} + \sqrt{\frac{V_{R,j}^*}{2\pi}}a_R^{\vphantom{\dag}}(k)c_j^{\dag} \right]dk + \int_{-\infty}^0\left[ \sqrt{\frac{V_{L,j}^{\vphantom{*}}}{2\pi}}a_L^{\dag}(k)c_j^{\vphantom{\dag}} + \sqrt{\frac{V_{L,j}^*}{2\pi}}a_L^{\vphantom{\dag}}(k)c_j^{\dag} \right]dk.
\label{eq_app:Hj_approx1}
\end{split}
\end{align}
\vspace{-0.2in}
\end{widetext}

The Hamiltonian $H_j$ commutes with 
\begin{align}
\begin{split}
N_j^{\vphantom{\dag}} =&\; \frac{1}{2}\sigma_{z,j}^{\vphantom{\dag}} + c_j^{\dag}c_j^{\vphantom{\dag}} + \int_0^{\infty}a_R^\dag(k)a_R^{\vphantom{\dag}}(k)dk \\
&+ \int_{-\infty}^0a_L^\dag(k)a_L^{\vphantom{\dag}}(k)dk,
\end{split}
\end{align}
the total excitation number operator for cavity $j$ (i.e., ${\comm{H_j}{N_j} = 0}$). This means that the total excitation number is a conserved quantity, which we can see because in $H_j$ every creation operator is paired with an annihilation operator, meaning that no physical process described by $H_j$ can change the total number of excitations. We can therefore shift the energy spectrum using the transformation ${H_j \rightarrow H_j - \omega_0 N_j}$, as this is a constant energy shift that does not affect the dynamics of the system. After using the transformation, we absorb remaining factors of $\omega_0$ into the definitions of the emitter and cavity frequencies (i.e., ${\omega_{e,j} - \omega_0 \rightarrow \omega_{e,j}}$, and ${\omega_{c,j} - \omega_0 \rightarrow \omega_{c,j}}$).

Next, we extend the integration limits in $H_j$ such that all the lower limits are ${k = -\infty}$ and all the upper limits are ${k = \infty}$. This is well-justified within the regime of validity of the linear dispersion approximation, where only photon wave numbers close to $k_0$ (for ${k>0}$) and $-k_0$ (for ${k<0}$) are considered. After extending the integration limits, the substitution ${k' = k-k_0}$ can be used in the integrals containing $a_R(k)$, and ${k' = k+k_0}$ can be used in the integrals containing $a_L(k)$. The new integration variable $k'$ can then be relabelled with $k$, leading to the expression in Eq.~(\ref{eq_app:Hj_approx2}) below.
\begin{widetext}
\vspace{-0.2in}
\begin{align}
\begin{split}
H_j^{\vphantom{\dag}} =&\; \frac{1}{2}\omega_{e,j}^{\vphantom{\dag}}\sigma_{z,j}^{\vphantom{\dag}} + \omega_{c,j}^{\vphantom{\dag}}c_j^{\dag}c_j^{\vphantom{\dag}} + \int_{-\infty}^{\infty} \! v_g k \;a_R^\dag(k+k_0)a_R^{\vphantom{\dag}}(k+k_0)dk - \int_{-\infty}^{\infty} \! v_g k \;a_L^\dag(k-k_0)a_L^{\vphantom{\dag}}(k-k_0)dk + g_j^{\vphantom{\dag}}\sigma_j^+c_j^{\vphantom{\dag}} + g_j^{*\vphantom{\dag}}\sigma_j^-c_j^{\dag}\\[0.05in]
&+ \int_{-\infty}^{\infty} \! \left[ \! \sqrt{\frac{V_{R,j}^{\vphantom{*}}}{2\pi}}a_R^{\dag}(k+k_0)c_j^{\vphantom{\dag}} + \sqrt{\frac{V_{R,j}^*}{2\pi}}a_R^{\vphantom{\dag}}(k+k_0)c_j^{\dag} \right] \! dk + \int_{-\infty}^{\infty} \! \left[ \! \sqrt{\frac{V_{L,j}^{\vphantom{*}}}{2\pi}}a_L^{\dag}(k-k_0)c_j^{\vphantom{\dag}} + \sqrt{\frac{V_{L,j}^*}{2\pi}}a_L^{\vphantom{\dag}}(k-k_0)c_j^{\dag} \right] \! dk.
\label{eq_app:Hj_approx2}
\end{split}
\end{align}
\vspace{-0.2in}
\end{widetext}

We complete the transformation from $k$-space to frequency-space by defining the frequency variable ${\omega = v_gk}$ for integrals containing ${a_R(k+k_0)}$ and ${\omega = -v_gk}$ for integrals containing ${a_L(k-k_0)}$, as well as the frequency-space operators ${a_R(\omega) = a_R(k+k_0)/\sqrt{v_g}}$ and ${a_L(\omega) = a_L(k-k_0)/\sqrt{v_g}}$. We also absorb remaining factors of the group velocity into the cavity-waveguide coupling rates, i.e., ${V_{R,j}/v_g \rightarrow V_{R,j}}$ and ${V_{L,j}/v_g \rightarrow V_{L,j}}$. The final frequency-space Hamiltonian is given in Eq.~(\ref{eq_app:Hj_freq}) below.
\begin{widetext}
\vspace{-0.2in}
\begin{align}
\begin{split}
H_j^{\vphantom{\dag}} =&\;\frac{1}{2}\omega_{e,j}^{\vphantom{\dag}}\sigma_{z,j}^{\vphantom{\dag}} + \omega_{c,j}^{\vphantom{\dag}}c_j^{\dag}c_j^{\vphantom{\dag}} + \int_{-\infty}^{\infty}\omega \left[ a_R^\dag(\omega)a_R^{\vphantom{\dag}}(\omega) + a_L^\dag(\omega)a_L^{\vphantom{\dag}}(\omega) \right]d\omega + g_j^{\vphantom{\dag}}\sigma_j^+c_j^{\vphantom{\dag}} + g_j^{*\vphantom{\dag}}\sigma_j^-c_j^{\dag} \\[0.05in]
&+ \int_{-\infty}^{\infty}\left[ \sqrt{\frac{V_{R,j}^{\vphantom{*}}}{2\pi}}a_R^{\dag}(\omega)c_j^{\vphantom{\dag}} + \sqrt{\frac{V_{R,j}^*}{2\pi}}a_R^{\vphantom{\dag}}(\omega)c_j^{\dag} + \sqrt{\frac{V_{L,j}^{\vphantom{*}}}{2\pi}}a_L^{\dag}(\omega)c_j^{\vphantom{\dag}} + \sqrt{\frac{V_{L,j}^*}{2\pi}}a_L^{\vphantom{\dag}}(\omega)c_j^{\dag} \right]d\omega.
\label{eq_app:Hj_freq}
\end{split}
\end{align}
\vspace{-0.15in}
\end{widetext}


\section{Transfer Matrix Derivations}\label{sec_app:B}


\subsection{Single-Cavity Transfer Matrices}\label{sec_app:B1}

Here we outline the derivation of the cavity transfer matrices $T_j$ [Eq.~(\ref{eq:Tj})], which involves using the frequency-space Hamiltonian $H_j$ in Eq.~(\ref{eq_app:Hj_freq}) to derive the input-output relations for the single-cavity system in Fig.~\ref{fig_app:single_cavity}. First, we derive the Heisenberg equations that describe the evolution of the waveguide operators ${a_R(\omega,t)}$ and ${a_L(\omega,t)}$ in time $t$:
\begin{align}
\begin{split}
\frac{d}{dt}a_{\mu}(\omega,t) &= i\comm{H_j^{\vphantom{\dag}}}{a_{\mu}(\omega,t)}\\[0.05in]
&= -i\omega a_{\mu}(\omega,t) - i \sqrt{\frac{V_{\mu,j}^{\vphantom{*}}}{2\pi}} c_j^{\vphantom{\dag}}(t),
\label{eq_app:wg_Heisenberg}
\end{split}
\end{align}
where ${\mu \in \{ L,R \}}$, and we used the bosonic commutation relations $\smash{\comm{a_{\mu}^{\vphantom{\dag}}(\omega,t)}{a_{\mu}^{\dag}(\omega',t)} = \delta(\omega-\omega')}$, with all other equal-time commutators involving the waveguide operators being zero. These commutators follow from the $k$-space commutators $\smash{\comm{a_{\mu}^{\vphantom{\dag}}(k,t)}{a_{\mu}^{\dag}(k',t)} = \delta(k-k')}$, and the definitions of the frequency-space waveguide operators from Appendix~\ref{sec_app:A}. We now multiply both sides of Eq.~(\ref{eq_app:wg_Heisenberg}) by $e^{i\omega t}$ and rearrange to obtain
\begin{equation}
\frac{d}{dt}\Bigl[ a_{\mu}(\omega,t) e^{i\omega t} \Bigr] = -i \sqrt{\frac{V_{\mu,j}^{\vphantom{*}}}{2\pi}} c_j^{\vphantom{\dag}}(t) e^{i\omega t}.
\label{eq_app:wg_Heisenberg_2}
\end{equation}
Relabelling $t$ with $t'$ and subsequently integrating from an `input time' $t_0$ to some time $t$ leads to
\begin{equation}
a_{\mu}(\omega,t) e^{i\omega t} - a_{\mu}(\omega,t_0) e^{i\omega t_0} = -i \sqrt{\frac{V_{\mu,j}^{\vphantom{*}}}{2\pi}} \int_{t_0}^t c_j^{\vphantom{\dag}}(t') e^{i\omega t'} dt'.
\end{equation}
Multiplying each term by $e^{-i\omega t}$ and then integrating over all $\omega$ gives
\begin{align}
\begin{split}
\int_{-\infty}^{\infty}&a_{\mu}(\omega,t) d\omega - \int_{-\infty}^{\infty}a_{\mu}(\omega,t_0) e^{-i\omega(t-t_0)} d\omega \\[0.05in]
&= -2 \pi i \sqrt{\frac{V_{\mu,j}^{\vphantom{*}}}{2\pi}} \int_{t_0}^t dt' c_j^{\vphantom{\dag}}(t') \left[ \int_{-\infty}^{\infty} \frac{d\omega}{2\pi} e^{i\omega(t'-t)} \right] \\[0.05in]
&= -2 \pi i \sqrt{\frac{V_{\mu,j}^{\vphantom{*}}}{2\pi}} \int_{t_0}^t dt' c_j^{\vphantom{\dag}}(t') \delta(t'-t) \\[0.05in]
&= -i\pi \sqrt{\frac{V_{\mu,j}^{\vphantom{*}}}{2\pi}} c_j^{\vphantom{\dag}}(t),
\end{split}
\end{align}
where the integral with respect to $t'$ gives a factor of ${1/2}$ because ${\delta(t'-t)}$ is centred at one of the integration limits. Dividing through by $\sqrt{2\pi}$ leads to
\begin{equation}
\frac{1}{\sqrt{2\pi}}\int_{-\infty}^{\infty}a_{\mu}(\omega,t) d\omega - a_{\mu,\text{in}}(t) = -\frac{i}{2} \sqrt{V_{\mu,j}^{\vphantom{*}}} c_j^{\vphantom{\dag}}(t),
\label{eq_app:input}
\end{equation}
where
\begin{equation}
a_{\mu,\text{in}}(t) = \frac{1}{\sqrt{2\pi}}\int_{-\infty}^{\infty}a_{\mu}(\omega,t_0) e^{-i\omega(t-t_0)} d\omega
\end{equation}
is the definition of an input operator in the input-output formalism~\cite{Gardiner1985}. Returning to Eq.~(\ref{eq_app:wg_Heisenberg_2}), relabelling $t$ with $t'$, integrating from some time $t$ to an `output time' $t_1$ and repeating the remaining steps gives
\begin{equation}
a_{\mu,\text{out}}(t) - \frac{1}{\sqrt{2\pi}}\int_{-\infty}^{\infty}a_{\mu}(\omega,t) d\omega = -\frac{i}{2} \sqrt{V_{\mu,j}^{\vphantom{*}}} c_j^{\vphantom{\dag}}(t),
\label{eq_app:output}
\end{equation}
where
\begin{equation}
a_{\mu,\text{out}}(t) = \frac{1}{\sqrt{2\pi}}\int_{-\infty}^{\infty}a_{\mu}(\omega,t_1) e^{-i\omega(t-t_1)} d\omega
\end{equation}
is the definition of an output operator in the input-output formalism~\cite{Gardiner1985}. Eqs.~(\ref{eq_app:input}) and (\ref{eq_app:output}) can easily be rearranged to obtain the following expressions for the input and output operators:
\begin{subequations}
\begin{align}
a_{\mu,\text{in}}(t) &= \frac{1}{\sqrt{2\pi}} \int_{-\infty}^{\infty} a_{\mu}(\omega,t) d\omega + \frac{i}{2}\sqrt{V_{\mu,j}^{\vphantom{*}}}c_j^{\vphantom{\dag}}(t),\label{eq_app:input2}\\[0.05in]
a_{\mu,\text{out}}(t) &= \frac{1}{\sqrt{2\pi}} \int_{-\infty}^{\infty} a_{\mu}(\omega,t) d\omega - \frac{i}{2}\sqrt{V_{\mu,j}^{\vphantom{*}}}c_j^{\vphantom{\dag}}(t),\label{eq_app:output2}
\end{align}
\end{subequations}
which immediately lead to the input-output relations
\begin{equation}
a_{\mu,\text{out}}(t) = a_{\mu,\text{in}}(t) - i\sqrt{V_{\mu,j}^{\vphantom{*}}}c_j^{\vphantom{\dag}}(t).
\label{eq_app:input_output_relations}
\end{equation}

From the input-output relations in Eq.~(\ref{eq_app:input_output_relations}), we can obtain the single-cavity transfer matrix $T_j$ that relates the input and output modes $a_{L,\text{in}}$, $a_{R,\text{out}}$  on the right side of cavity $j$ to the input and output modes $a_{R,\text{in}}$, $a_{L,\text{out}}$ on the left side of the cavity (see Fig.~\ref{fig_app:single_cavity}) by eliminating the cavity operator $c_j(t)$. This can be achieved using the Heisenberg equation for $c_j(t)$, given in Eq.~(\ref{eq_app:cavity_Heisenberg}) below. We derive this again using the frequency-space Hamiltonian $H_j$ from Eq.~(\ref{eq_app:Hj_freq}), and we use the bosonic commutation relations for the cavity mode operators, i.e., $\smash{\bigl[ c_j^{\vphantom{\dag}}(t),c_j^{\dag}(t) \bigr] = 1}$ (all other equal-time commutators involving the cavity operators are zero). In Eq.~(\ref{eq_app:cavity_Heisenberg}), we also use the definitions of the input operators $a_{R,\text{in}}(t)$ and $a_{L,\text{in}}(t)$ from Eq.~(\ref{eq_app:input2}) to eliminate the integrals.
\begin{widetext}
\begin{align}
\begin{split}
\frac{d}{dt}c_j^{\vphantom{\dag}}(t) &= i\comm{H_j^{\vphantom{\dag}}}{c_j^{\vphantom{\dag}}(t)} = -i\omega_{c,j}^{\vphantom{\dag}}c_j^{\vphantom{\dag}}(t) - ig_j^{*\vphantom{\dag}}\sigma_j^-(t) - i\sqrt{\frac{V_{R,j}^*}{2\pi}} \int_{-\infty}^{\infty} a_R^{\vphantom{\dag}}(\omega,t) d\omega - i\sqrt{\frac{V_{L,j}^*}{2\pi}} \int_{-\infty}^{\infty} a_L^{\vphantom{\dag}}(\omega,t) d\omega \\[0.05in]
&= \biggl( -i\omega_{c,j}^{\vphantom{\dag}} - \frac{1}{2}|V_{R,j}^{\vphantom{\dag}}| - \frac{1}{2}|V_{L,j}^{\vphantom{\dag}}| \biggr) c_j^{\vphantom{\dag}}(t) - ig_j^{*\vphantom{\dag}}\sigma_j^-(t) - i\biggl[ \sqrt{V_{R,j}^*}a_{R,\text{in}}^{\vphantom{\dag}}(t) + \sqrt{V_{L,j}^*}a_{L,\text{in}}^{\vphantom{\dag}}(t)\biggr].
\label{eq_app:cavity_Heisenberg}
\end{split}
\end{align}
\end{widetext}
If we assume coherent driving at frequency $\omega$, the evolution of the cavity operator can be approximated as ${c_j(t) \approx c_j(0)e^{-i\omega t}}$, provided that the cavity couples to the driving field more strongly than to the two-level emitter within it. This means that $\smash{\frac{d}{dt}c_j(t) \approx -i\omega c_j(t)}$ which, upon substitution into Eq.~(\ref{eq_app:cavity_Heisenberg}), yields an algebraic equation involving $c_j(t)$, i.e.,
\begin{align}
\begin{split}
\biggl(-i&\Delta_{c,j}^{\vphantom{\dag}} + \frac{1}{2}|V_{R,j}^{\vphantom{\dag}}| + \frac{1}{2}|V_{L,j}^{\vphantom{\dag}}| \biggr) c_j^{\vphantom{\dag}}(t) + ig_j^{*\vphantom{\dag}}\sigma_j^-(t) \\[0.05in]
&= - i\biggl[ \sqrt{V_{R,j}^*}a_{R,\text{in}}^{\vphantom{\dag}}(t) + \sqrt{V_{L,j}^*}a_{L,\text{in}}^{\vphantom{\dag}}(t)\biggr],
\label{eq_app:cavity_Heisenberg_approx}
\end{split}
\end{align}
where ${\Delta_{c,j} = \omega - \omega_{c,j}}$ is the frequency detuning between the input coherent driving field and cavity $j$. This equation can be solved for $c_j(t)$ in terms of the input operators $a_{R,\text{in}}(t)$ and $a_{L,\text{in}}(t)$ once the lowering operator $\sigma_j^-(t)$ of emitter $j$ is eliminated. The Heisenberg equation for $\sigma_j^-(t)$ is
\begin{align}
\begin{split}
\frac{d}{dt}\sigma_j^-(t) &= i\comm{H_j^{\vphantom{\dag}}}{\sigma_j^-(t)} \\[0.05in]
&= -i\omega_{e,j}^{\vphantom{\dag}}\sigma_j^-(t) + ig_j^{\vphantom{\dag}}\sigma_{z,j}^{\vphantom{\dag}}(t)c_j^{\vphantom{\dag}}(t),
\label{eq_app:emitter_Heisenberg}
\end{split}
\end{align}
for which we again use the Hamiltonian from Eq.~(\ref{eq_app:Hj_freq}), and the operator definitions ${\sigma_j^+(0) = \ketbra{e_j}{g_j}}$, ${\sigma_j^-(0) = \ketbra{g_j}{e_j}}$, and ${\sigma_{z,j}(0) = \ketbra{e_j}{e_j} - \ketbra{g_j}{g_j}}$. Again, by approximating the time evolution as being dominated by the coherent driving field with frequency $\omega$, we have ${\sigma_j^-(t) \approx \sigma_j^-(0)e^{-i\omega t}}$, and hence $\smash{\frac{d}{dt}\sigma_j^-(t) \approx -i\omega \sigma_j^-(t)}$. Furthermore, assuming that the coherent input is weak leads to the simplification ${\sigma_{z,j}(t) \approx -1}$ for all $t$, which is the weak-excitation approximation~\cite{Rephaeli2012}. With these approximations, the Heisenberg equation for $\sigma_j^-(t)$ in Eq.~(\ref{eq_app:emitter_Heisenberg}) reduces to
\begin{equation}
\sigma_j^-(t) = \frac{g_j}{\Delta_{e,j}} c_j^{\vphantom{\dag}}(t),
\end{equation}
where ${\Delta_{e,j} = \omega - \omega_{e,j}}$ is the frequency detuning between the input coherent driving field and emitter $j$. Substituting this result into Eq.~(\ref{eq_app:cavity_Heisenberg_approx}) and then rearranging for $c_j(t)$ gives
\begin{equation}
c_j^{\vphantom{\dag}}(t) = \frac{\sqrt{V_{R,j}^*}a_{R,\text{in}}^{\vphantom{\dag}}(t) + \sqrt{V_{L,j}^*}a_{L,\text{in}}^{\vphantom{\dag}}(t)}{\Delta_{c,j} - \frac{|g_j|^2}{\Delta_{e,j}} + \frac{i}{2}\bigl( |V_{R,j}| + |V_{L,j}| \bigr)}.
\end{equation}
Finally, the above expression for $c_j(t)$ can be substituted into the input-output relations in Eq.~(\ref{eq_app:input_output_relations}), which leads to simultaneous equations that relate the input and output modes in the single-cavity subsystem shown in Fig.~\ref{fig_app:single_cavity}. These can be written in the matrix equation
\begin{equation}
\begin{pmatrix}
    a_{R,\text{out}} \\
    a_{L,\text{in}}
\end{pmatrix}
= T_j
\begin{pmatrix}
    a_{R,\text{in}} \\
    a_{L,\text{out}}
\end{pmatrix},
\end{equation}
where $T_j$ is the transfer matrix for cavity $j$, given in Eq.~(\ref{eq_app:Tj}) below and in Eq.~(\ref{eq:Tj}) in terms of $\smash{\alpha_j^{(\pm)} = \frac{i}{2}\left( |V_{R,j}| \pm |V_{L,j}| \right)}$, $\smash{\beta_j = \Delta_{c,j} - |g_j|^2/\Delta_{e,j}}$, and ${\zeta_j = -i\left( V_{R,j}^{\vphantom{*}} V_{L,j}^* \right)^{\frac{1}{2}}}$.
\begin{widetext}
\begin{equation}
T_j = \frac{1}{\Delta_{c,j} - \frac{|g_j|^2}{\Delta_{e,j}} + \frac{i}{2}\bigl( |V_{R,j}| - |V_{L,j}| \bigr)}
\begin{pmatrix}
    \Delta_{c,j} - \frac{|g_j|^2}{\Delta_{e,j}} - \frac{i}{2}\bigl( |V_{R,j}| + |V_{L,j}| \bigr) & -i\sqrt{V_{R,j}^{\phantom{*}}V_{L,j}^*} \\[0.15in]
    i\sqrt{V_{L,j}^{\phantom{*}}V_{R,j}^*} & \Delta_{c,j} - \frac{|g_j|^2}{\Delta_{e,j}} + \frac{i}{2}\bigl( |V_{R,j}| + |V_{L,j}| \bigr)
\end{pmatrix}.
\label{eq_app:Tj}
\end{equation}
\end{widetext}

In the derivation of $T_j$ we assumed a weak coherent input field, corresponding to photons of a single frequency $\omega$. However, in our work we apply this result in calculations involving wave packets with a finite spectral width. This is valid because different frequency components are independent when we neglect multi-photon nonlinearities such as four-wave mixing.


\subsection{Waveguide Transfer Matrices}\label{sec_app:B2}

We now show how the waveguide transfer matrices $T_{\text{wg}}^{(i,j)}$ in Eq.~(\ref{eq:Twg}) can be derived. In the waveguide regions of length $d_{i,j}$ separating the cavities, photon propagation is governed by the free waveguide Hamiltonian
\begin{equation}
H_{\text{wg}} = \int_{-\infty}^{\infty}\omega \left[ a_R^\dag(\omega)a_R^{\vphantom{\dag}}(\omega) + a_L^\dag(\omega)a_L^{\vphantom{\dag}}(\omega) \right]d\omega,
\end{equation}
which is based on the linear dispersion approximation [free waveguide term in Eq.~(\ref{eq_app:Hj_freq})], as shown in Appendix~\ref{sec_app:A}. The Heisenberg equations for the waveguide operators ${a_R(\omega,t)}$ and ${a_L(\omega,t)}$ in this case are therefore
\begin{equation}
\frac{d}{dt}a_{\mu}(\omega,t) = i\bigl[H_{\text{wg}}, a_{\mu}(\omega,t)\bigr] = -i\omega a_{\mu}(\omega,t),
\end{equation}
where again ${\mu \in \{L,R \}}$. These have the trivial solutions ${a_{\mu}(\omega,t) = a_{\mu}(\omega,0) e^{-i\omega t}}$. For photon propagation between neighbouring cavities $(i,j)$ over the distance $d_{i,j}$, the evolution occurs for a time ${t = d_{i,j}/v_g}$, so we define the input operators ${a_{\mu,\text{in}} = a_{\mu}(\omega,0)}$ at ${t=0}$ and the output operators ${a_{\mu,\text{out}} = a_{\mu}(\omega,d_{i,j}/v_g)}$ at ${t=d_{i,j}/v_g}$. With these definitions, the relationships between the output modes and the input modes for a waveguide region of length $d_{i,j}$ are
\begin{equation}
a_{\mu,\text{out}} = a_{\mu,\text{in}} e^{-i\omega d_{i,j}/v_g}.
\end{equation}
These can be written in matrix form as 
\begin{equation}
\begin{pmatrix}
    a_{R,\text{out}} \\
    a_{L,\text{in}}
\end{pmatrix}
= T_{\text{wg}}^{(i,j)}
\begin{pmatrix}
    a_{R,\text{in}} \\
    a_{L,\text{out}}
\end{pmatrix},
\end{equation}
where
\begin{equation}
T_{\text{wg}}^{(i,j)} = 
\begin{pmatrix}
    e^{-i\omega d_{i,j}/v_g} & 0 \\[0.05in]
    0 & e^{i\omega d_{i,j}/v_g}
\end{pmatrix}
\label{eq_app:Twg}
\end{equation}
is the transfer matrix that describes free photon propagation over the distances $d_{i,j}$ in the waveguide, as given in Eq.~(\ref{eq:Twg}).


\section{Cavity and Emitter Losses}\label{sec_app:C}

In any physical realisation of the waveguide-coupled cavities, photons will leak out from the cavities into the environment and emitters will have a non-zero probability of coupling to non-cavity modes, resulting in photon loss from the system. We can include losses from cavity $j$ and emitter $j$ in the relevant transfer matrix $T_j$ by adding Lindblad terms to the Heisenberg equations for the cavity operator $c_j$ and the emitter operator $\sigma_j^-$. In particular, the Lindblad operator $\smash{L_{c,j} = \sqrt{\kappa_j}c_j}$ describes photon loss from cavity $j$ at rate $\kappa_j$, and the Lindblad operator $\smash{L_{e,j} = \sqrt{\gamma_j}\sigma_j^-}$ describes photon loss from emitter $j$ at rate $\gamma_j$. For an operator $A(t)$ corresponding to an observable in a system with Hamiltonian $H$, the Lindblad master equation in the Heisenberg picture is
\begin{equation}
\frac{d}{dt}A(t) = i\bigl[H, A(t)\bigr] + \sum_k \left[ L_k^{\dag} A(t) L_k^{\vphantom{\dag}} - \frac{1}{2}\Bigl\{L_k^{\dag}L_k^{\vphantom{\dag}},A(t)\Bigr\} \right],
\end{equation}
where the $L_k$ are Lindblad operators. Including Lindblad terms with the operators $L_{c,j}$ and $L_{e,j}$ in the Heisenberg equations for $c_j(t)$ and $\sigma_j^-(t)$ given in Eqs.~(\ref{eq_app:cavity_Heisenberg}) and (\ref{eq_app:emitter_Heisenberg}) leads to Eqs.~(\ref{eq_app:cavity_losses}) and (\ref{eq_app:emitter_losses}) below.
\begin{widetext}
\begin{subequations}
\begin{align}
\begin{split}
\frac{d}{dt}c_j^{\vphantom{\dag}}(t) &= i\comm{H_j^{\vphantom{\dag}}}{c_j^{\vphantom{\dag}}(t)} + L_{c,j}^{\dag}c_j^{\vphantom{\dag}}(t)L_{c,j}^{\vphantom{\dag}} - \frac{1}{2}\Bigl\{L_{c,j}^{\dag}L_{c,j}^{\vphantom{\dag}},c_j^{\vphantom{\dag}}(t)\Bigr\} \\[0.05in]
&= -i\left( \omega_{c,j}^{\vphantom{\dag}} - \frac{i\kappa_j}{2}\right)c_j^{\vphantom{\dag}}(t) - ig_j^{*\vphantom{\dag}}\sigma_j^-(t) - i\sqrt{\frac{V_{R,j}^*}{2\pi}} \int_{-\infty}^{\infty} a_R^{\vphantom{\dag}}(\omega,t) d\omega - i\sqrt{\frac{V_{L,j}^*}{2\pi}} \int_{-\infty}^{\infty} a_L^{\vphantom{\dag}}(\omega,t) d\omega,
\label{eq_app:cavity_losses}
\end{split}
\end{align}
\begin{align}
\begin{split}
\frac{d}{dt}\sigma_j^-(t) &= i\comm{H_j^{\vphantom{\dag}}}{\sigma_j^-(t)} + L_{e,j}^{\dag}\sigma_j^-(t)L_{e,j}^{\vphantom{\dag}} - \frac{1}{2}\Bigl\{L_{e,j}^{\dag}L_{e,j}^{\vphantom{\dag}},\sigma_j^-(t)\Bigr\} \\[0.05in]
&= -i\left( \omega_{e,j}^{\vphantom{\dag}} - \frac{i\gamma_j}{2}\right)\sigma_j^-(t) + ig_j^{\vphantom{\dag}}\sigma_{z,j}^{\vphantom{\dag}}(t)c_j^{\vphantom{\dag}}(t).
\label{eq_app:emitter_losses}
\end{split}
\end{align}
\end{subequations}
\end{widetext}

Comparing Eqs.~(\ref{eq_app:cavity_losses}) and (\ref{eq_app:emitter_losses}) with Eqs.~(\ref{eq_app:cavity_Heisenberg}) and (\ref{eq_app:emitter_Heisenberg}), we see that introducing the Lindblad terms simply amounts to using the substitutions ${\omega_{c,j} \rightarrow \omega_{c,j} - i\kappa_j/2}$ and ${\omega_{e,j} \rightarrow \omega_{e,j} - i\gamma_j/2}$ in the original Heisenberg equations for $c_j(t)$ and $\sigma_j^-(t)$~\cite{Rephaeli2013}. In terms of the frequency detunings ${\Delta_{c,j} = \omega - \omega_{c,j}}$ and ${\Delta_{e,j} = \omega - \omega_{e,j}}$, these substitutions are equivalent to ${\Delta_{c,j} \rightarrow \Delta_{c,j} + i\kappa_j/2}$ and ${\Delta_{e,j} \rightarrow \Delta_{e,j} + i\gamma_j/2}$. Since these substitutions do not modify any subsequent steps in the calculation of $T_j$ in Appendix~\ref{sec_app:B}, they can be used directly in the final result in Eq.~(\ref{eq_app:Tj}). This allows us to include losses from all the cavities and emitters in our system within the transmission and reflection coefficients that we calculate using the transfer matrices.


\section{Analytical Results for Identical, Equally-Spaced Cavities}\label{sec_app:D}

In general, each cavity can have a different resonance frequency $\omega_{c,j}$, each emitter can have a different transition frequency $\omega_{e,j}$, the cavity separations $d_{i,j}$ can all be different, and both the coupling rates $g_j$, $V_{R,j}$, $V_{L,j}$ and loss rates $\kappa_j$, $\gamma_j$ can vary from cavity to cavity. In this general situation, the parameter space increases in size as the number of cavities $N$ increases, since every additional cavity brings an extra eight parameters to the model. This makes obtaining analytical results for transmission and reflection more challenging for larger $N$. In this appendix, we consider the ideal case where all the cavities are identical and equally-spaced. Here, the number of parameters in the model is independent of $N$. To simplify the general waveguide-coupled cavity system shown in Fig.~\ref{fig:N_cavities_diagram} to this special case, we set ${\omega_{c,j} = \omega_c}$, ${\omega_{e,j} = \omega_e}$, ${g_j = g}$, ${V_{R,j} = V_R}$, ${V_{L,j} = V_L}$, ${\kappa_j = \kappa}$, and ${\gamma_j = \gamma}$ for all ${j \in \{1,2,\dotsc,N\}}$, and ${d_{i,j} = d}$ for all ${(i,j) \in \{(1,2), (2,3), \dotsc, (N-1,N)\}}$. We neglect cavity and emitter losses in the calculation that follows, and include them in the final results using the substitutions described in Appendix~\ref{sec_app:C}. A consequence of the above simplifications is that all the cavity transfer matrices $T_j$ are identical to each other, and all the waveguide transfer matrices $\smash{T_{\text{wg}}^{(i,j)}}$ are identical to each other. If we write ${T_j = T_c}$ for all the cavities and $\smash{T_{\text{wg}}^{(i,j)} = T_{\text{wg}}}$ for all the free photon propagation regions, from Eq.~(\ref{eq:Ttot}) it follows that
\begin{equation}
T_{\text{tot}} = \bigl(T_c \hspace{0.025in} T_{\text{wg}}\bigr)^N
\label{eq_app:Ttot_identical}
\end{equation}
is the total transfer matrix, where we have multiplied by an additional waveguide transfer matrix $T_{\text{wg}}$ from the right, which will simplify the rest of the calculation. Physically, this corresponds to offsetting the input/output photon phase by a constant factor and hence only introduces a global, unobservable phase $e^{\pm i\omega d/v_g}$ that does not affect the transmission $|t_N|^2$ and reflection $|r_N|^2$.

Eq.~(\ref{eq_app:Ttot_identical}) shows that, in order to calculate the total transfer matrix $T_{\text{tot}}$ (and hence the transmission and reflection coefficients), we only need to find the $N$th power of the product matrix ${T_c\hspace{0.025in}T_{\text{wg}}}$. This product is given in Eq.~(\ref{eq_app:TcTwg}), which we find using Eqs.~(\ref{eq_app:Tj}) and (\ref{eq_app:Twg}). In Eq.~(\ref{eq_app:TcTwg}), ${\Delta_c = \omega - \omega_c}$ (${\Delta_e = \omega - \omega_e}$) is the frequency detuning between the input photons and the cavities (emitters).
\begin{widetext}
\begin{align}
\begin{split}
T_c \hspace{0.025in} T_{\text{wg}} &= \frac{1}{\Delta_c - \frac{|g|^2}{\Delta_e} + \frac{i}{2}\bigl( |V_R| - |V_L| \bigr)}
\begin{pmatrix}
    \Delta_c - \frac{|g|^2}{\Delta_e} - \frac{i}{2}\bigl( |V_R| + |V_L| \bigr) & -i\sqrt{V_R^{\vphantom{*}} V_L^*} \\[0.1in]
    i\sqrt{V_L^{\vphantom{*}} V_R^*} & \Delta_c - \frac{|g|^2}{\Delta_e} + \frac{i}{2}\bigl( |V_R| + |V_L| \bigr)
\end{pmatrix}
\begin{pmatrix}
    \color{white}\dot{\color{black}e^{-i\omega d/v_g}} & 0 \\[0.1in]
    0 & \color{white}\dot{\color{black}e^{i\omega d/v_g}}
\end{pmatrix}\\[0.25in]
&= \frac{1}{\Delta_c - \frac{|g|^2}{\Delta_e} + \frac{i}{2}\bigl( |V_R| - |V_L| \bigr)}
\begin{pmatrix}
    \Bigl[ \Delta_c - \frac{|g|^2}{\Delta_e} - \frac{i}{2}\bigl( |V_R| + |V_L| \bigr) \Bigr] e^{-i\omega d/v_g} & -i e^{i\omega d/v_g} \sqrt{V_R^{\vphantom{*}} V_L^*} \\[0.1in]
    i e^{-i\omega d/v_g} \sqrt{V_L^{\vphantom{*}} V_R^*} & \Bigl[ \Delta_c - \frac{|g|^2}{\Delta_e} + \frac{i}{2}\bigl( |V_R| + |V_L| \bigr) \Bigr] e^{i\omega d/v_g}
\end{pmatrix}.
\label{eq_app:TcTwg}
\end{split}
\end{align}
\vspace{0.2in}
\end{widetext}
The total transfer matrix for $N$ identical, equally-spaced waveguide-coupled cavities therefore has the form
\begin{equation}
T_{\text{tot}} = \frac{1}{\Bigl[\Delta_c - \frac{|g|^2}{\Delta_e} + \frac{i}{2}\bigl( |V_R| - |V_L| \bigr)\Bigr]^N}A^N,
\label{eq_app:Ttot_AN}
\end{equation}
where $A$ is the final ${2 \times 2}$ matrix in Eq.~(\ref{eq_app:TcTwg}) without the prefactor, which has the simple form
\begin{equation}
A = 
\begin{pmatrix}
    a & b \\
    b^* & a^*
\end{pmatrix},
\label{eq_app:A_matrix}
\end{equation}
with
\begin{subequations}
\begin{gather}
a = \biggl[ \Delta_c - \frac{|g|^2}{\Delta_e} - \frac{i}{2}\bigl( |V_R| + |V_L| \bigr) \biggr] e^{-i\omega d/v_g},\label{eq_app:a}\\[0.1in]
b = -i e^{i\omega d/v_g} \sqrt{V_R^{\vphantom{*}} V_L^*}.\label{eq_app:b}
\end{gather}
\end{subequations}
We can obtain $A^N$ by diagonalising the matrix $A$. From the characteristic equation ${|A - \lambda_{\pm} \mathbbm{1}| = 0}$ (where $\mathbbm{1}$ is the identity matrix and $|\dotsc|$ denotes the determinant), we find the eigenvalues of $A$ to be
\begin{equation}
\lambda_\pm^{\vphantom{N}} = \frac{1}{2} (a + a^*) \pm  \frac{1}{2} \sqrt{ (a - a^*)^2 + 4|b|^2 },
\label{eq_app:eigenvalues1}
\end{equation}
and from the eigenvalue equations ${A \boldsymbol{u}_\pm = \lambda_\pm \boldsymbol{u}_\pm}$, it follows that the eigenvectors are
\begin{equation}
\boldsymbol{u}_\pm^{\vphantom{N}} = C_\pm^{\vphantom{N}}
\begin{pmatrix}
    b \\
    \lambda_\pm^{\vphantom{N}} - a
\end{pmatrix},
\end{equation}
where $C_\pm$ are normalisation constants, which ensure that ${\boldsymbol{u}_\pm \cdot \boldsymbol{u}_\pm = 1}$. By matrix diagonalisation, we therefore have ${A = PDP^{-1}}$, where
\begin{equation}
D = 
\begin{pmatrix}
    \lambda_+^{\vphantom{N}} & 0 \\
    0 & \lambda_-^{\vphantom{N}}
\end{pmatrix}
\vspace{0.1in}
\end{equation}
is the diagonal matrix that contains the eigenvalues of $A$, and
\begin{equation}
P = 
\begin{pmatrix}
    C_+^{\vphantom{N}} b & C_-^{\vphantom{N}} b \\[0.05in]
    C_+^{\vphantom{N}} \bigl(\lambda_+^{\vphantom{N}} - a\bigr) \hspace{0.025in} & \hspace{0.025in} C_-^{\vphantom{N}} \bigl(\lambda_-^{\vphantom{N}} - a\bigr)
\end{pmatrix}
\end{equation}
is the matrix constructed from the normalised eigenvectors of $A$. It then follows that ${A^N = P D^N P^{-1}}$, where $D^N$ is simply the matrix $D$ with the diagonal elements $\lambda_\pm$ raised to the power $N$. The result obtained for $A^N$ after calculating the inverse matrix $P^{-1}$ and performing the matrix multiplication is given in Eq.~(\ref{eq_app:AN_matrix}).
\begin{widetext}
\begin{align}
\begin{split}
A^N = P D^N P^{-1} &= 
\begin{pmatrix}
    C_+^{\vphantom{N}} b & C_-^{\vphantom{N}} b \\[0.05in]
    C_+^{\vphantom{N}} \bigl(\lambda_+^{\vphantom{N}} - a\bigr) \hspace{0.025in} & \hspace{0.025in} C_-^{\vphantom{N}} \bigl(\lambda_-^{\vphantom{N}} - a\bigr)
\end{pmatrix}
\begin{pmatrix}
    \lambda_+^N & 0 \\[0.05in]
    0 & \lambda_-^N
\end{pmatrix}
\begin{pmatrix}
    C_+^{\vphantom{N}} b & C_-^{\vphantom{N}} b \\[0.05in]
    C_+^{\vphantom{N}} \bigl(\lambda_+^{\vphantom{N}} - a\bigr) \hspace{0.025in} & \hspace{0.025in} C_-^{\vphantom{N}} \bigl(\lambda_-^{\vphantom{N}} - a\bigr)
\end{pmatrix}^{-1}\\[0.2in]
&= \frac{1}{\lambda_-^{\vphantom{N}} - \lambda_+^{\vphantom{N}}}
\begin{pmatrix}
    \lambda_+^N \bigl(\lambda_-^{\vphantom{N}} - a\bigr) - \lambda_-^N \bigl(\lambda_+^{\vphantom{N}} - a\bigr)\quad & b \bigl(\lambda_-^N - \lambda_+^N\bigr)\\[0.1in]
    \frac{1}{b} \bigl(\lambda_+^N - \lambda_-^N\bigr) \bigl(\lambda_+^{\vphantom{N}} - a\bigr) \bigl(\lambda_-^{\vphantom{N}} - a\bigr)\quad & \lambda_-^N \bigl(\lambda_-^{\vphantom{N}} - a\bigr) - \lambda_+^N \bigl(\lambda_+^{\vphantom{N}} - a\bigr)
\end{pmatrix}.
\label{eq_app:AN_matrix}
\end{split}
\end{align}
\end{widetext}
We can now substitute the result in Eq.~(\ref{eq_app:AN_matrix}) into the total transfer matrix $T_{\text{tot}}$ in Eq.~(\ref{eq_app:Ttot_AN}), and use Eq.~(\ref{eq:tN_and_rN}) to find the $N$-cavity transmission and reflection coefficients $t_N$ and $r_N$ from the matrix elements of $T_{\text{tot}}$. The final results for $t_N$ and $r_N$ that are valid for an arbitrary number $N$ of identical, equally-spaced cavities are given in Eqs.~(\ref{eq_app:tN_identical}) and (\ref{eq_app:rN_identical}) below in terms of the eigenvalues $\lambda_\pm$, which are given in terms of the system parameters in Eq.~(\ref{eq_app:eigenvalues2}) [obtained by substituting $a$ and $b$ from Eqs.~(\ref{eq_app:a}) and (\ref{eq_app:b}) into Eq.~(\ref{eq_app:eigenvalues1})]. The \mbox{oscillatory} transmission behaviour that we observe for ${N>1}$ due to interference in the waveguide is clearly visible in these analytical results. To include losses from the cavities and emitters, we can use ${\Delta_c \rightarrow \Delta_c + i\kappa/2}$ and ${\Delta_e \rightarrow \Delta_e + i\gamma/2}$ in Eqs.~(\ref{eq_app:tN_identical}), (\ref{eq_app:rN_identical}), and (\ref{eq_app:eigenvalues2}), as discussed in Appendix~\ref{sec_app:C}. We note that, when we use these substitutions, the matrix $A$ no longer has the form given in Eq.~(\ref{eq_app:A_matrix}), as the frequency detunings effectively become complex and the diagonal elements of $A$ are no longer complex conjugates of each other.
\begin{widetext}
\begin{subequations}
\begin{align}
t_N &= \frac{ \bigl( \lambda_+^{\vphantom{N}} - \lambda_-^{\vphantom{N}} \bigr) \left[ \Delta_c - \frac{|g|^2}{\Delta_e} - \frac{i}{2}\bigl( |V_R| - |V_L| \bigr) \right]^N }{ \bigl( \lambda_+^{N+1} - \lambda_-^{N+1} \bigr) - \bigl( \lambda_+^N - \lambda_-^N \bigr) \left[ \Delta_c - \frac{|g|^2}{\Delta_e} - \frac{i}{2}\bigl( |V_R| + |V_L| \bigr) \right] e^{-i\omega d/v_g} }, \label{eq_app:tN_identical} \\[0.15in]
r_N &= \frac{ -i e^{-i\omega d/v_g} \sqrt{V_L^{\vphantom{*}}V_R^*} \bigl(\lambda_+^N - \lambda_-^N\bigr) }{ \bigl(\lambda_+^{N+1} - \lambda_-^{N+1}\bigr) - \bigl(\lambda_+^N - \lambda_-^N\bigr) \Bigl[ \Delta_c - \frac{|g|^2}{\Delta_e} - \frac{i}{2}\bigl(|V_R|+|V_L|\bigr) \Bigr] e^{-i\omega d/v_g} }, \label{eq_app:rN_identical}
\end{align}
\end{subequations}
where
\begin{align}
\begin{split}
\hspace{0.12in}\lambda_\pm^{\vphantom{N}} =\;& \left(\Delta_c - \frac{|g|^2}{\Delta_e}\right) \cos\left(\frac{\omega d}{v_g}\right) - \frac{1}{2} \bigl(|V_R|+|V_L|\bigr) \sin\left(\frac{\omega d}{v_g}\right) \\[0.1in]
&\pm \sqrt{|V_R V_L| - \left[ \left(\Delta_c - \frac{|g|^2}{\Delta_e}\right) \sin\left(\frac{\omega d}{v_g}\right) + \frac{1}{2} \bigl(|V_R|+|V_L|\bigr) \cos\left(\frac{\omega d}{v_g}\right)\right]^2}.
\label{eq_app:eigenvalues2}
\end{split}
\end{align}
\end{widetext}

As previously mentioned, including the extra waveguide transfer matrix $T_{\text{wg}}$ in $T_{\text{tot}}$ [Eq.~(\ref{eq_app:Ttot_identical})] only gives rise to a global phase in $t_N$ and $r_N$, which does not affect the transmission and reflection spectra that we calculate using these results. However, since the global phase is $\omega$-dependent, it has observable consequences in the fidelities $F_{\nu}$ because we have to integrate over $\omega$ [it essentially becomes a relative phase in this case, see Eq.~(\ref{eq:F})]. This is why we use the general approach for calculating $t_N$ and $r_N$ outlined in the main body of the paper when we compute the switch fidelities, rather than the analytical results presented in this appendix.


\section{Analysis of Disorder in the Switch}\label{sec_app:E}

In this appendix, we analyse how disorder in the waveguide-coupled cavities could affect the transmission spectrum of the proposed switch, in the situation where fabrication imperfections cannot be overcome completely. In particular, we generate a Gaussian distribution of a chosen parameter, and observe how disorder in this parameter changes the ideal ten-cavity spectra in Fig.~\ref{fig:switch_10_cavities}. We consider disorder in the cavity mode wavelengths $\lambda_{c,j}$, the emitter wavelengths $\lambda_{e,j}$, the coupled-cavity $Q$ factors $Q_{c,j}$, and the nearest-neighbour cavity separations $d_{i,j}$. For each of these parameters, we generate a Gaussian distribution with a chosen \mbox{standard} deviation $\sigma$. After generating the parameters, we use Eq.~(\ref{eq:Ttot}) with ${N=10}$ to find the total transfer matrix for ten cavities, and obtain the transmission coefficient $t_{10}$ from the matrix elements using Eq.~(\ref{eq:tN_and_rN}). We then compare the transmission spectra in the disordered cases to the ideal case shown in Fig.~\ref{fig:switch_10_cavities}, in both the weak and strong emitter-cavity coupling regimes. After studying disorder in the parameters mentioned above, we consider how the operation of the switch is affected when the strong coupling regime cannot be reached in one of the cavities in the array.

\begin{figure*}
    \includegraphics[width = 1.45\columnwidth]{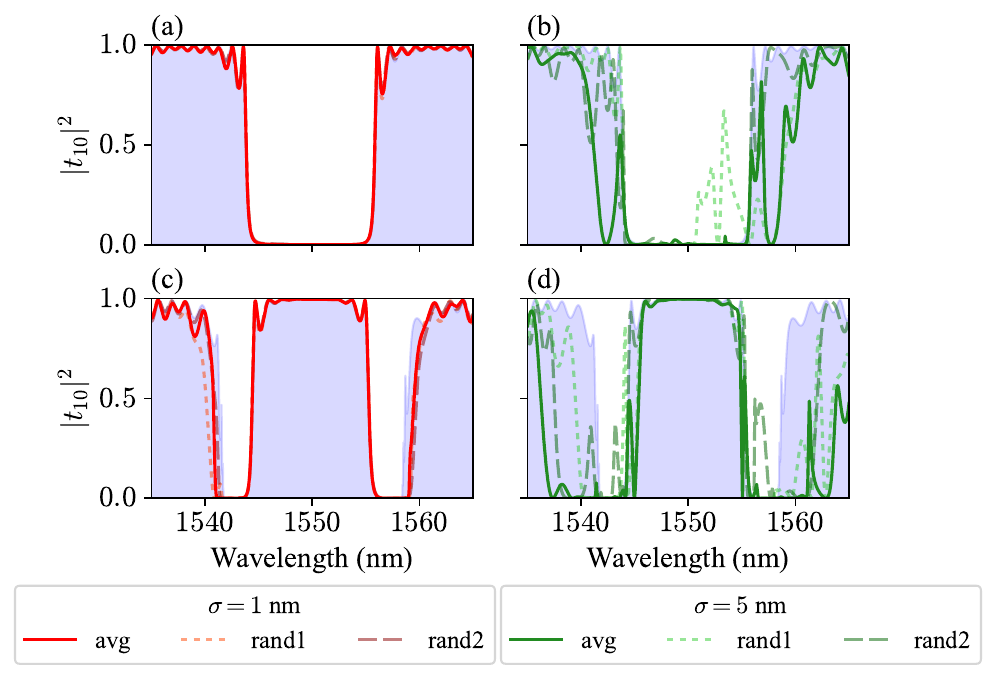}
    \caption{Transmission spectra for ten waveguide-coupled cavities, comparing the ideal case with no disorder from Fig.~\ref{fig:switch_10_cavities} (indicated by the shaded regions here) with the disordered case where the cavity mode wavelengths $\lambda_{c,j}$ form a Gaussian distribution with a mean of $1550$~nm and a standard deviation (a), (c) ${\sigma = 1}$~nm (red curves), or (b), (d) ${\sigma = 5}$~nm (green curves). (a), (b) correspond to the weak coupling regime (${g_j/2\pi = 100}$~MHz for all $j$) and (c), (d) correspond to the strong coupling regime ($g_j/2\pi = 1$~THz for all $j$).}
    \label{fig_app:cavity_disorder}
\end{figure*}

\begin{figure*}
    \includegraphics[width = 1.45\columnwidth]{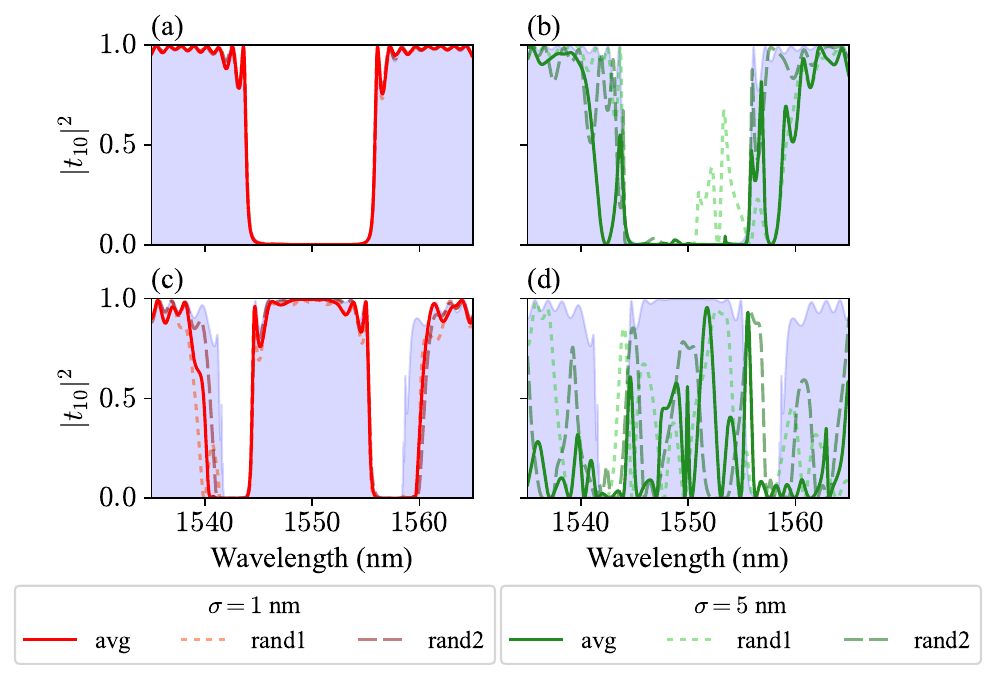}
    \caption{Same as Fig.~\ref{fig_app:cavity_disorder}, except that the emitters are now tuned on resonance with the disordered cavities (${\lambda_{e,j} = \lambda_{c,j}}$ for all $j$), instead of having identical transition wavelengths at the centre of the switching region.}
    \label{fig_app:cavity_emitter_disorder}
\end{figure*}

\begin{figure*}
    \includegraphics[width = 1.45\columnwidth]{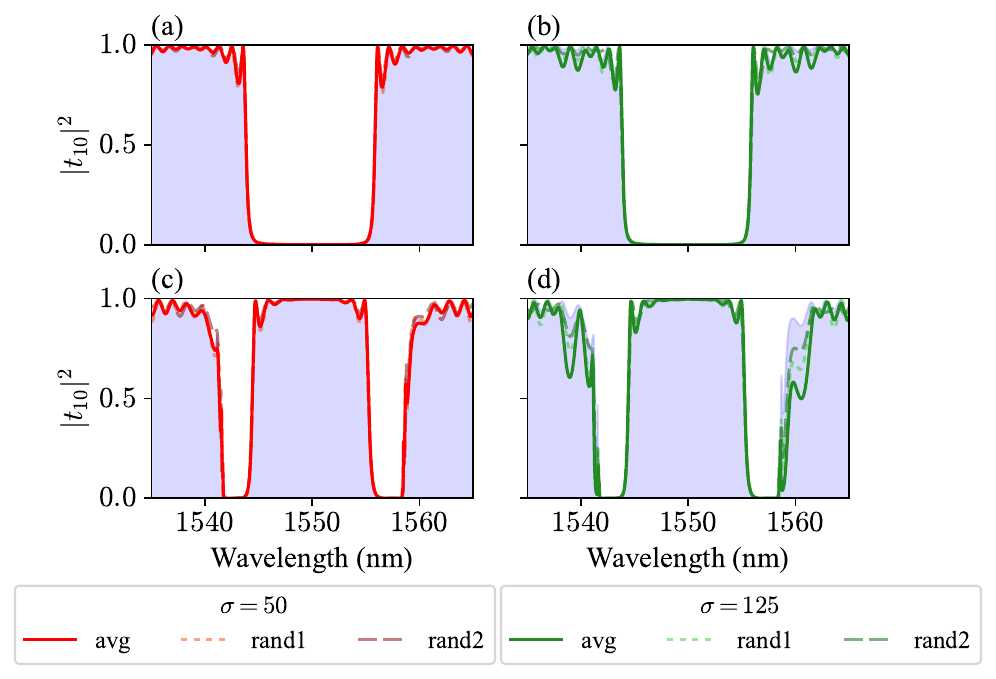}
    \caption{Transmission spectra for ten waveguide-coupled cavities, comparing the ideal case with no disorder (shaded regions) with the disordered case where the coupled-cavity $Q$ factors $Q_{c,j}$ form a Gaussian distribution with a mean of $500$ and a standard deviation (a), (c) ${\sigma = 50}$, i.e., $10\%$ of the mean (red curves), or (b), (d) ${\sigma = 125}$, i.e., $25\%$ of the mean (green curves). As in the previous figures in this appendix, (a), (b) correspond to the weak coupling regime (${g_j/2\pi = 100}$~MHz for all $j$) and (c), (d) correspond to the strong coupling regime (${g_j/2\pi = 1}$~THz for all $j$).}
    \label{fig_app:Qc_disorder}
\end{figure*}

\begin{figure*}
    \includegraphics[width = 1.45\columnwidth]{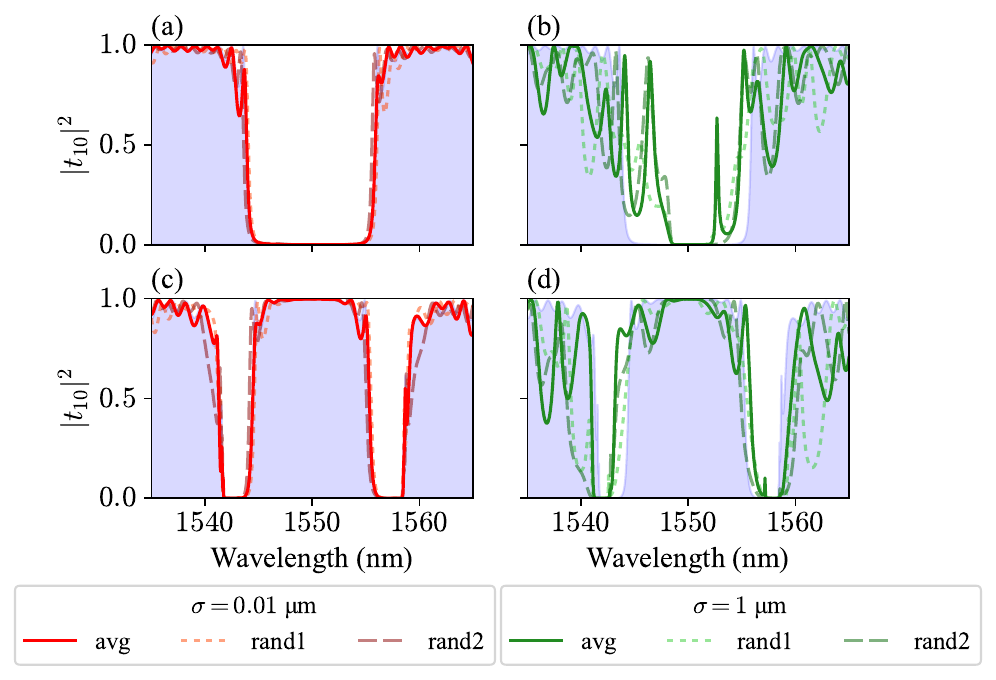}
    \caption{Transmission spectra for ten waveguide-coupled cavities, comparing the ideal case with no disorder (shaded regions) with the disordered case where the cavity separations $d_{i,j}$ form a Gaussian distribution with a mean of $31.5$~$\upmu$m and a standard deviation (a), (c) ${\sigma = 0.01}$~$\upmu$m (red curves), or (b), (d) ${\sigma = 1}$~$\upmu$m (green curves). (a), (b) correspond to the weak coupling regime (${g_j/2\pi = 100}$~MHz for all $j$) and (c), (d) correspond to the strong coupling regime (${g_j/2\pi = 1}$~THz for all $j$).}
    \label{fig_app:separation_disorder}
\end{figure*}

The various types of disorder are shown in Figs.~\ref{fig_app:cavity_disorder}-\ref{fig_app:separation_disorder}. In each of the figures, `avg' refers to a Gaussian distribution of a chosen disordered parameter that was obtained by averaging over $1000$ randomly generated Gaussian distributions with standard deviation $\sigma$, which shows the expected behaviour of the system with the specified standard deviation (solid curves). In addition, `rand1' and `rand2' correspond to two randomly generated, non-averaged Gaussian distributions that provide additional examples of how the switching behaviour may be affected by disorder for a given $\sigma$ (dashed curves). The ideal ten-cavity transmission spectra from Fig.~\ref{fig:switch_10_cavities} are indicated with shaded regions in each of these figures for ease of comparison.

We first consider disorder in cavity mode wavelengths $\lambda_{c,j}$ in Fig.~\ref{fig_app:cavity_disorder}, with a mean of $1550$~nm and a standard deviation ${\sigma = 1}$~nm [red curves, Figs.~\ref{fig_app:cavity_disorder}(a) and (c)] or ${\sigma = 5}$~nm [green curves, Figs.~\ref{fig_app:cavity_disorder}(b) and (d)]. All other parameters are the same as those used to generate the ideal spectra in Fig.~\ref{fig:switch_10_cavities} [${Q_{c,j} = 500}$, ${Q_{u,j} = 5 \times 10^4}$, ${\lambda_{e,j} = 1550}$~nm, and ${\gamma_j/2\pi = 1}$~GHz for all ${j \in \{1,2,\dotsc,10\}}$, ${d_{i,j} = 31.5}$~$\upmu$m for ${(i,j) \in \{(1,2), (2,3), \dotsc, (9,10)\}}$, and ${v_g = 0.3c}$]. Figs.~\ref{fig_app:cavity_disorder}(a) and (b) correspond to the weak emitter-cavity coupling regime (${g_j/2\pi = 100}$~MHz), and Figs.~\ref{fig_app:cavity_disorder}(c) and (d) correspond to the strong coupling regime (${g_j/2\pi = 1}$~THz). When ${\sigma = 1}$~nm, the FWHM [${2\sqrt{2\ln(2)}\sigma}$] of the Gaussian distribution of the cavity mode wavelengths $\lambda_{c,j}$ is less than the bandwidth of operation of the ideal switch (which is approximately $10$~nm, extending from about $1545$~nm to $1555$~nm, see shaded regions). Here, the change in transmission compared to the ideal case is small in the switching region, in both the weak [Fig.~\ref{fig_app:cavity_disorder}(a)] and strong [Fig.~\ref{fig_app:cavity_disorder}(c)] coupling regimes. Conversely, when ${\sigma = 5}$~nm, the FWHM of the cavity wavelength distribution is larger than the switching bandwidth, and we see that the transmission dip in the weak coupling regime can split into multiple narrower dips, resulting in undesired transmission features within the switching region [Fig.~\ref{fig_app:cavity_disorder}(b)]. The transmission window in the strong coupling regime [Fig.~\ref{fig_app:cavity_disorder}(d)] is also distorted to a greater degree than in Fig.~\ref{fig_app:cavity_disorder}(c). Based on these results, we expect that the performance of the switch will remain high as long as the distribution of the disordered cavity mode wavelengths does not exceed the switching bandwidth.

In Fig.~\ref{fig_app:cavity_emitter_disorder}, we show how the switch performs when the emitters are tuned on resonance with cavities that have disordered mode wavelengths $\lambda_{c,j}$. The only change compared to Fig.~\ref{fig_app:cavity_disorder} is that we now have ${\lambda_{e,j} = \lambda_{c,j}}$ for all $j$, rather than having all the emitters tuned to the centre of the switching region at $1550$~nm. As expected, Figs.~\ref{fig_app:cavity_emitter_disorder}(a) and (b) are identical to Figs.~\ref{fig_app:cavity_disorder}(a) and (b) respectively, as the emitter wavelengths are insignificant in the weak emitter-cavity coupling regime, where the presence of the emitters does not affect the transmission spectra. In the strong coupling regime, we observe that there is little change to the ideal switching bandwidth when the disorder in the emitter wavelengths is less than this bandwidth [${\sigma = 1}$~nm, Fig.~\ref{fig_app:cavity_emitter_disorder}(c)]. However, when ${\sigma = 5}$~nm [Fig.~\ref{fig_app:cavity_emitter_disorder}(d)], the FWHM of the wavelength distribution exceeds the width of the switching region, and the transmission of the switch can be completely destroyed. Therefore, for the Rabi splitting to produce the desired transmission window, the disorder in the emitter wavelengths cannot exceed the switching bandwidth.

Fig.~\ref{fig_app:Qc_disorder} shows how disorder in the coupled-cavity $Q$ factors $Q_{c,j}$ affects the switching operation. We consider a Gaussian distribution with a mean coupled-$Q$ factor of $500$ and a standard deviation ${\sigma = 50}$, i.e., $10\%$ of the mean value [red curves, Figs.~\ref{fig_app:Qc_disorder}(a) and (c)], or ${\sigma = 125}$, i.e., $25\%$ of the mean value [green curves, Figs.~\ref{fig_app:Qc_disorder}(b) and (d)]. All other parameters are the same as those used for ten identical cavities in Fig.~\ref{fig:switch_10_cavities} [in particular, ${\lambda_{c,j} = \lambda_{e,j} = 1550}$~nm for all $j$, and ${d_{i,j} = 31.5}$~$\upmu$m for all nearest neighbours $(i,j)$]. We see that for both ${\sigma = 50}$ and ${\sigma = 125}$ the transmission in the switching region remains almost identical to the ideal case indicated by the shaded regions, implying that the proposed switch is robust against reasonably large variations in the cavity $Q$ factors, and hence against variations in the cavity-waveguide coupling rates ${V_{R,j} = V_{L,j} = \omega_{c,j}/2Q_{c,j}}$.

Next, we consider a Gaussian distribution of nearest-neighbour cavity separations $d_{i,j}$ in Fig.~\ref{fig_app:separation_disorder}, with a mean of $31.5$~$\upmu$m and a standard deviation ${\sigma = 0.01}$~$\upmu$m [red curves, Figs.~\ref{fig_app:separation_disorder}(a) and (c)] or ${\sigma = 1}$~$\upmu$m [green curves, Figs.~\ref{fig_app:separation_disorder}(b) and (d)]. All other parameters are identical to those used for ten cavities in Fig.~\ref{fig:switch_10_cavities} (${\lambda_{c,j} = \lambda_{e,j} = 1550}$~nm and ${Q_{c,j} = 500}$ for all $j$). We observe that, for sub-wavelength disorder on the order of $10$~nm [Figs.~\ref{fig_app:separation_disorder}(a) and (c)], the change compared to the ideal spectra is very small, while for wavelength-scale disorder on the order of $1$~$\upmu$m [Figs.~\ref{fig_app:separation_disorder}(b) and (d)], the transmission spectrum can be significantly distorted. This is caused by modified interference resulting from a change in the phase shifts acquired by photons in the waveguide due to the modified cavity separations.

\begin{figure}
    \includegraphics[width = 0.8\columnwidth]{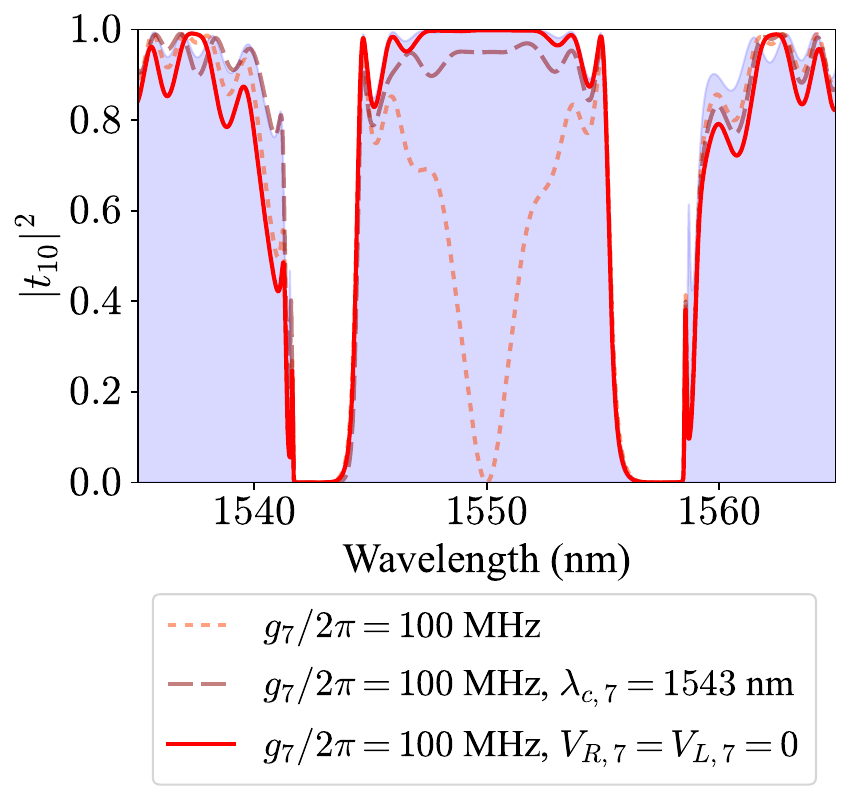}
    \caption{Transmission spectra for ten waveguide-coupled cavities, comparing the ideal strong coupling case with no disorder [shaded region, same as Fig.~\ref{fig:switch_10_cavities}(b)] with the case where emitter $7$ is weakly coupled to its cavity, i.e., ${g_7/2\pi = 100}$~MHz, ${g_j/2\pi = 1}$~THz otherwise (light, dashed red curve). We also show how the high transmission window is recovered when the cavity with the weakly coupled emitter is either detuned to ${\lambda_{c,7} = 1543}$~nm (dark, dashed red curve), or decoupled from the waveguide completely by setting ${V_{R,7} = V_{L,7} = 0}$ (solid red curve).}
    \label{fig_app:g_disorder}
\end{figure}

Finally, we consider how the transmission spectrum of the switch is affected if one cavity in the array does not satisfy the strong coupling condition when the switch is operated in transmission mode. In particular, in Fig.~\ref{fig_app:g_disorder}, we consider the situation where ${g_j/2\pi = 1}$~THz for all $j$ except ${j=7}$, where we have ${g_7/2\pi = 100}$~MHz. In this way, we take into account the situation where not all emitters may be strongly coupled to their cavities even if they are tuned on resonance, for example due to the positional dependence of the emitter-cavity coupling arising from the spatial profile of the cavity mode fields. While we choose emitter $7$ to be weakly coupled in this example, we note that a different emitter choice would not significantly change the transmission spectra shown in Fig.~\ref{fig_app:g_disorder}. The transmission in the system of ten waveguide-coupled cavities with emitter $7$ weakly coupled and all the others being strongly coupled is shown by the light, dashed red curve in Fig.~\ref{fig_app:g_disorder}, where we leave all other parameters identical to those used in Fig.~\ref{fig:switch_10_cavities}(b) for ten cavities in the strong coupling regime (shaded region here). We see that the high transmission window changes into a transmission dip at the cavity resonance wavelength, which arises because we effectively obtain a convolution of the spectrum corresponding to nine waveguide-coupled cavities containing strongly coupled emitters with the spectrum of a single cavity with a weakly coupled emitter [as in Fig.~\ref{fig:switch_1_cavity}(a)]. This result implies that all emitters must be strongly coupled to their cavities for the switching operation to work as intended. Fortunately, there are several \mbox{different} approaches we can take to recover the high transmission window if there is a `bad' cavity that cannot satisfy the strong coupling condition. One option is to detune the cavity containing the weakly coupled emitter away from the switching bandwidth (in the present example, this would be cavity $7$). We show this with the dark, dashed red curve in Fig.~\ref{fig_app:g_disorder}, for which we only change the resonance wavelength of the bad cavity with the weakly coupled emitter from $1550$~nm to ${\lambda_{c,7} = 1543}$~nm compared to the light, dashed red curve. We see that detuning the bad cavity recovers the high transmission window. The further away the cavity is detuned, the closer the transmission in the switching region becomes to the ideal case indicated with the shading. Another option would be to decouple the bad cavity from the waveguide completely, for example by physically displacing it away from the waveguide to reduce the spatial overlap between the waveguide and cavity modes (this is the same mechanism that we can use to modulate the cavity $Q$ factors). We show this with the solid red curve in Fig.~\ref{fig_app:g_disorder} by setting ${V_{R,7} = V_{L,7} = 0}$ for the coupling rates between cavity $7$ and the waveguide (all other parameters are the same as for the light, dashed red curve). We see that decoupling the bad cavity in this way returns the transmission window very close to the ideal case. Regardless of whether we detune the cavity from the switching region or decouple it completely from the rest of the system, we effectively end up with a switch operating with nine waveguide-coupled cavities. This does not reduce the transmission bandwidth because the bandwidth in the strong coupling regime is determined by the Rabi splitting of the cavity modes. The transmission spectrum is more sensitive to the cavity number in the weak coupling regime (i.e., in reflection mode), where it is preferable to have all the cavities coupled to the waveguide and tuned to the same resonance wavelength.


\section{Efficiency and Fidelity Derivations}\label{sec_app:F}

In the final appendix, we derive the efficiency and fidelity expressions given in Eqs.~(\ref{eq:E}) and (\ref{eq:F}). We consider the right-moving single-photon input wave packet
\begin{equation}
\ket{\psi_{\text{in}}} = \int_{-\infty}^{\infty} d\omega f(\omega) a_{R,\text{in}}^{\dag}(\omega) \ket{0},
\end{equation}
with Gaussian envelope
\begin{equation}
f(\omega) = \left[ \frac{4\ln(2)}{\pi \sigma_{\omega}^2} \right]^{1/4} e^{-2\ln(2)(\omega - \omega_{\text{cen}})^2/\sigma_{\omega}^2},
\end{equation}
where $\omega_{\text{cen}}$ is the central frequency and $\sigma_{\omega}$ is the FWHM in frequency units. This satisfies the normalisation condition
\begin{equation}
\braket{\psi_{\text{in}}}{\psi_{\text{in}}} = \int_{-\infty}^{\infty} \left|f(\omega)\right|^2 d\omega = 1.
\end{equation}
After the interaction with the cavities takes place, the output state will in general be a superposition of transmitted and \mbox{reflected} components:
\begin{align}
\begin{split}
\ket{\psi_{\text{out}}} = \int_{-\infty}^{\infty} d\omega &\left[ t_N(\omega)f(\omega) a_{R,\text{out}}^{\dag}(\omega) \right.\\
&\left. + r_N(\omega)f(\omega) a_{L,\text{out}}^{\dag}(\omega) \right] \ket{0}.
\end{split}
\end{align}

When the switch is operating in reflection mode (weak emitter-cavity coupling regime), the ideal left-moving output wave packet would be
\begin{equation}
\ket{\psi_{\text{r,id}}} = \int_{-\infty}^{\infty} d\omega f(\omega) a_{L,\text{out}}^{\dag}(\omega) \ket{0},
\end{equation}
while the actual (unnormalised) reflected wave packet is
\begin{equation}
\ket{\psi_{\text{r,act}}} = \int_{-\infty}^{\infty} d\omega \; r_N(\omega) f(\omega) a_{L,\text{out}}^{\dag}(\omega) \ket{0}.
\end{equation}
We define the reflection fidelity $F_r$ as the modulus-squared of the overlap between the actual and ideal reflected wave packets,
\begin{equation}
F_r = \left| \braket{\psi_{\text{r,id}}}{\psi_{\text{r,act}}} \right|^2 = \left| \int_{-\infty}^{\infty} r_N(\omega) \left|f(\omega)\right|^2 d\omega \right|^2,
\end{equation}
and we define the reflection efficiency $E_r$ as the modulus-squared of the overlap between the actual reflected wave packet and itself, which is equal to one if the input wave packet is guaranteed to be reflected (i.e., ${|r_N(\omega)|^2 = 1}$ for all $\omega$), or less than one if there is a non-zero probability of the wave packet being transmitted or lost into the environment:
\begin{equation}
E_r = \left| \braket{\psi_{\text{r,act}}}{\psi_{\text{r,act}}} \right|^2 = \left| \int_{-\infty}^{\infty} \left|r_N(\omega)\right|^2 \left|f(\omega)\right|^2 d\omega \right|^2.
\end{equation}

When the switch is operating in transmission mode (strong emitter-cavity coupling regime), the ideal right-moving output wave packet is
\begin{equation}
\ket{\psi_{\text{t,id}}} = \int_{-\infty}^{\infty} d\omega f(\omega) a_{R,\text{out}}^{\dag}(\omega) \ket{0},
\end{equation}
while the actual (unnormalised) transmitted wave packet is
\begin{equation}
\ket{\psi_{\text{t,act}}} = \int_{-\infty}^{\infty} d\omega \; t_N(\omega) f(\omega) a_{R,\text{out}}^{\dag}(\omega) \ket{0}.
\end{equation}
Analogously to the reflection case, we define the transmission fidelity $F_t$ as the modulus-squared of the overlap between the actual and ideal transmitted wave packets,
\begin{equation}
F_t = \left| \braket{\psi_{\text{t,id}}}{\psi_{\text{t,act}}} \right|^2 = \left| \int_{-\infty}^{\infty} t_N(\omega) \left|f(\omega)\right|^2 d\omega \right|^2,
\end{equation}
and we define the transmission efficiency $E_t$ as the modulus-squared of the overlap between the actual transmitted wave packet and itself:
\begin{equation}
E_t = \left| \braket{\psi_{\text{t,act}}}{\psi_{\text{t,act}}} \right|^2 = \left| \int_{-\infty}^{\infty} \left|t_N(\omega)\right|^2 \left|f(\omega)\right|^2 d\omega \right|^2.
\end{equation}

In order to obtain the results presented in the main body of the paper, we express the efficiencies and fidelities in terms of wavelength $\lambda$. Using ${\omega = 2\pi c/\lambda}$, ${d\omega = -(2\pi c/\lambda^2) d\lambda}$, and ${\sigma_{\omega} = (2\pi c/\lambda_{\text{cen}}^2) \sigma_{\lambda}}$, we arrive at
\begin{equation}
E_{\nu} = \left| \int_{-\infty}^{\infty} \left|\tilde{\nu}_N(\lambda)\right|^2 |\tilde{f}(\lambda)|^2 d\lambda \right|^2
\label{eq_app:E_wavelength}
\end{equation}
for the efficiencies and 
\begin{equation}
F_{\nu} = \left| \int_{-\infty}^{\infty} \tilde{\nu}_N(\lambda) |\tilde{f}(\lambda)|^2 d\lambda \right|^2
\label{eq_app:F_wavelength}
\end{equation}
for the fidelities (${\nu \in \{r,t\}}$), where ${\tilde{\nu}_N(\lambda) = \nu_N(2\pi c/\lambda)}$, and
\begin{equation}
\tilde{f}(\lambda) = \frac{\lambda_{\text{cen}}}{\lambda} \left[ \frac{4\ln(2)}{\pi \sigma_{\lambda}^2} \right]^{1/4} e^{-2\ln(2)(\lambda_{\text{cen}}/\lambda)^2(\lambda - \lambda_{\text{cen}})^2/\sigma_{\lambda}^2}
\end{equation}
is the transformed Gaussian wave packet, with central wavelength $\lambda_{\text{cen}}$ (${\omega_{\text{cen}} = 2\pi c/\lambda_{\text{cen}}}$) and FWHM $\sigma_{\lambda}$ in wavelength units. We note that the relationship ${\sigma_{\omega} = (2\pi c/\lambda_{\text{cen}}^2) \sigma_{\lambda}}$ between the FWHM in frequency and wavelength units is only valid for narrow wave packets, where ${\sigma_{\lambda} \ll \lambda_{\text{cen}}}$. Since we consider input wave packets with central wavelength ${\lambda_{\text{cen}} = 1550}$~nm and values of $\sigma_{\lambda}$ up to $1$~nm, we satisfy this condition. In addition, we can restrict the range of integration in Eqs.~(\ref{eq_app:E_wavelength}) and (\ref{eq_app:F_wavelength}) to a few nanometres around $\lambda_{\text{cen}}$ to a very good approximation, as the value of $\tilde{f}(\lambda)$ is negligible outside of this range for wave packet widths up to $1$~nm.


%

\end{document}